\documentclass[aps,prb,reprint,superscriptaddress,letterpaper]{revtex4-1}
\usepackage[english]{babel}
\usepackage{ucs}
\usepackage[utf8x]{inputenc}
\usepackage{amsmath}
\usepackage{amsfonts}
\usepackage{amssymb}
\usepackage{graphicx}
\usepackage{bm}
\usepackage{bbm}
\usepackage{empheq}

\newcommand{\bra}[1]{\langle #1|}
\newcommand{\ket}[1]{|#1\rangle}

\newcommand{\mm}[1]{\mathrm{#1}}
\newcommand{\ui}{\mathrm{i}}
\newcommand{\ue}{\mathrm{e}}
\newcommand{\ud}{\mathrm{d}}

\newcommand{\abs}[1]{\left|#1\right|}

\newcommand{\id}[1]{\mathrm{d} #1\,}

\newcommand{\Sv}{\mbox{\boldmath$S$}}
\newcommand{\hv}{\mbox{\boldmath$h$}}
\newcommand{\Iv}{\mbox{\boldmath$I$}}
\newcommand{\rv}{\mbox{\boldmath$r$}}
\newcommand{\qv}{\mbox{\boldmath$q$}}
\newcommand{\rvs}{\mbox{\boldmath$\scriptstyle{r}$}}
\newcommand{\qvs}{\mbox{\boldmath$\scriptstyle{q}$}}
\newcommand{\qvss}{\mbox{\boldmath$\scriptscriptstyle{q}$}}
\newcommand{\mps}{\mbox{$\scriptstyle{\mp}$}}
\newcommand{\Bv}{\mbox{\boldmath$B$}}

\begin{document}

\title{Interplay of Charge and Spin Coherence in
  Landau-Zener-Stückelberg-Majorana Interferometry}

\author{Hugo Ribeiro}
\thanks{Current address: Department of Physics, University of Basel, Klingelbergstrasse 82, CH-4056 Basel, Switzerland}
\affiliation{Department of Physics, University of Konstanz, D-78457 Konstanz, Germany}

\author{J. R. Petta}
\affiliation{Department of Physics, Princeton University, Princeton, New Jersey 08544, USA}
\affiliation{Princeton Institute for the Science and Technology of Materials (PRISM), Princeton University, Princeton, New Jersey 08544, USA}

\author{Guido Burkard}
\affiliation{Department of Physics, University of Konstanz, D-78457 Konstanz, Germany}



\begin{abstract}
We study Landau-Zener dynamics in a double quantum dot filled with two electrons, where
the spin states can become correlated with charge states, and the level velocity can be
tuned in a time-dependent fashion. We show that a correct interpretation of experimental
data is only possible when finite-time effects are taking into account.  In addition, our
formalism allows the study of partial adiabatic dynamics in the presence of
phonon-mediated hyperfine relaxation and charge noise induced dephasing.  Our findings
demonstrate that charge noise severely impacts the visibility of LZSM interference
fringes. This indicates that charge coherence must be treated on an equal footing with
spin coherence.
\end{abstract}

\maketitle

\section{Introduction}

Electron spins trapped in quantum dots (QDs) are promising candidates for implementing a
scalable quantum computer~\cite{loss1998, hanson2007}. Most of the DiVincenzo criteria,
which state physical requirements a system must fulfill to achieve quantum computing,
have been achieved with spin qubits, including initialization~\cite{ono2002},
readout~\cite{elzerman2004}, and coherent
control~\cite{petta2005,koppens2006,nowack2007}. While experimental progress has been
impressive, many of these methods are not yet accurate enough to allow large scale
quantum computing with single spins. A key challenge can be appreciated by considering
the relevant timescales associated with the spin dynamics. In GaAs quantum dot devices,
it is well known that the hyperfine interaction leads to a randomly fluctuating nuclear
field, $B_{\rm n}$ $\sim 2$ mT, which results in a $10$--$20$ ns inhomogeneous spin dephasing
time~\cite{petta2005,coish2005,taylor2007}. For comparison, the Rabi period obtained in a
GaAs double quantum dot (DQD) using conventional electron spin resonance (ESR) is on the
order of $110$ ns~\cite{koppens2006}. The use of spin-orbit driven electric dipole spin
resonance (EDSR) in GaAs leads to even slower Rabi periods of roughly $210$
ns~\cite{nowack2007}. Therefore, in the case of single spin rotations, gate operation
times are nearly an order of magnitude slower than the inhomogeneous spin dephasing time.
This issue is not specific to GaAs based nanostructures. In InAs nanowires, where the
spin-orbit interaction is larger than in GaAs, a Rabi period of $\sim
17\,\mm{ns}$~\cite{nadjperge2010} was
reported. However, it still is twice as long as the spin dephasing time of $\sim
8\,\mm{ns}$.

Viewed from a different perspective, the maximum ac field generated in DQD ESR
experiments is on the order of $2$ mT, which is the same magnitude as the fluctuating
nuclear field~\cite{koppens2006}. As a result, single spin rotations in GaAs qubits
follow imperfect trajectories on the Bloch sphere, resulting in a reduced oscillation
visibility and gate errors~\cite{koppens2006}. Single spin selectivity imposes an
additional challenge upon the development of a spin-based quantum processor. Magnetic
fields generated in ESR are difficult to localize on the nanometer scale. Without
g-factor control, or local magnetic field gradients, the spins located in a quantum
register would rotate at the same rate in the presence of a global ESR field. The long
term goal is to be able to drive selective single spin rotations, without affecting
neighboring spins that are on average only $20$--$50$ nm away.

Instead of using the spin-up and spin-down states of a single electron, the qubit basis
states can be represented by two (out of four) two-electron spin states confined in a
DQD~\cite{levy2002}. For a qubit whose basis states are encoded in the singlet $\mm{S}$
and triplet $\mm{T}_0$ spin states of a DQD, the two-electron exchange interaction allows
for fast single-qubit gates (hundreds of picoseconds)~\cite{petta2005}.  Recent
experiments~\cite{weperen2011,shulman2012} have also demonstrated the possibility of
realizing a conditional two-qubit gate. Within this two-spin version of a qubit, the
two-qubit gate realized in Ref.~\onlinecite{petta2005} can be interpreted as a single
qubit operation. However the exchange interaction only allows for rotations about a
single axis, whereas to generate arbitrary rotations one needs two perpendicular rotation
axes.  The generation of a nuclear magnetic field gradient~\cite{foletti2009} provides
rotations about a second, non-collinear axis. While remarkable, this method also presents
some difficulties when it has to be extended to a large number of qubits. It requires
that the nuclear polarization is controlled in each DQD to create the desired gradient
field. An advantage of this method is that the generation of the nuclear field gradient
reduces nuclear spin fluctuations, resulting in an increase in the spin dephasing
time~\cite{khaetskii2002}.

Recently, it has been proposed to use a two-spin basis consisting of the singlet $\mm{S}$
and triplet $\mm{T}_+$ spin states~\cite{petta2010,ribeiro2010,ribeiro2012}. Quantum
control of the $\mm{S}$-$\mm{T}_+$ qubit relies on
Landau-Zener-Stückelberg-Majorana~\cite{landau1932,zener1932,stuckelberg1932,majorana1932}
(LZSM) physics, which occurs in the system when the $\mm{S}$-$\mm{T}_+$ qubit is
repeatedly swept through the hyperfine mediated $\mm{S}$-$\mm{T}_+$ anti-crossing. This
all-electrical method also has the advantage of addressing individually each quantum dot.

LZSM physics describing the passage of a two-level quantum system through an
anti-crossing can be applied to different fields of physics and
chemistry~\cite{shevchenko2010}. In quantum information science, LZSM theory describes
accurately the observed interference fringes (Stückelberg oscillations) of a
superconducting qubit driven back and forth through its anti-crossing~\cite{oliver2005}.
The LZSM description also accurately describes the coherent manipulation of a two-spin
qubit encoded in the $\mm{S}$ and triplet $\mm{T}_+$ spin states, with dynamics driven by
repeated passages through a hyperfine-mediated
anti-crossing~\cite{petta2010,ribeiro2010}. In self-assembled quantum dots, LZSM theory
has been used to design high-fidelity all-optical control of spin-based
qubits~\cite{calarco2003,gauger2008,cole2008}.

In this article we develop a quantum master equation to describe partial adiabatic
passages between the spin singlet $\mm{S}$ and triplet $\mm{T}_+$ states in the presence
of both the fluctuating Overhauser field and the fluctuating charge environment. With our
theory, we show charge dynamics can significantly hinder LZSM interferometry of spin
states. While most of the interesting spin dynamics happens in the $(1,1)$ charge
configuration, with one electron per dot, initialization and measurements are done deep
in the $(2,0)$ charge configuration, where both electrons are in the left dot. Here
$(l,r)$ denotes the number of electrons in the left and right dot. Crossing the $(1,1)$
$\leftrightarrow$ $(2,0)$ interdot charge transition necessarily involves charge
dynamics. Since superpositions of charge states dephase on shorter time scales than
superpositions of spin states, it is essential to consider this fast effective
decoherence mechanism when spin and charge degrees of freedom become correlated during
spin qubit evolution~\cite{hayashi2003,petta2004}. In particular, we use our formalism to
analyze spin-charge dynamics associated with detuning pulses that have a tunable level
velocity, with a high level velocity away from the anti-crossing and a slow level
velocity in the vicinity of the anti-crossing~\cite{ribeiro2012}.

The paper is organized as follows. We start by reviewing standard Landau-Zener
theory, which is valid for an infinitely long ramp through an anti-crossing with a
constant level velocity. In Sec.~\ref{sec:adiabatic} we review the solution of the
finite-time LZSM model, which can be used to model realistic experiments, and we
demonstrate within the scope of this theory how fine tuning of the level velocity can be
used to increase the visibility of the quantum oscillations. Section~\ref{sec:st+}
focuses on the physical implementation of LZSM physics in a two-electron spin qubit. We
derive an effective Hamiltonian describing the dynamics of the states in the vicinity of
the $\mm{S}$-$\mm{T}_+$ anti-crossing. Compared to previously derived effective
Hamiltonians~\cite{ribeiro2010}, we include the effects of charge superposition states in the
spin-dependent anti-crossing. The last part of the section is devoted to the derivation
of a master equation that describes the evolution of the density matrix. In
Sec.~\ref{sec:results}, we first compare solutions of the master equation obtained with
exprimental pulse profiles and measurements performed on a GaAs double quantum dot.  We
then show theory results for the singlet return probability for which we explore the
effects of charge induced decoherence.

\section{Adiabatic control of a quantum two-level system}
\label{sec:adiabatic}

There are numerous problems in physics that deal with the physics of two-level systems.
The most common example is Rabi's formula~\cite{rabi1937} which describes the occupation
of a two-level system that is driven by a coherent field. In quantum information science,
Rabi oscillations are widely used, e.g., to manipulate an electron spin confined in a
QD~\cite{hanson2008}. Another widely studied problem involving only two quantum levels is
adiabatic passage, which is commonly employed in nuclear magnetic
resonance~\cite{abragam1983}. The physics of adiabatic passage can be found in a variety
of systems, and several theoretical models have been developed to describe different
kinds of adiabatic
processes~\cite{rosen1932,allen1987,hioe1984,bambini1981,demkov1969,hioe1985,zakrzewski1985,demkov1964,nikitin1962,carroll1986}.
There is, however, a particular description that has proven to be applicable in many
distinct fields of physics: the Landau-Zener~\cite{landau1932,zener1932} model. We refer
to it in this article as the LZSM model, since it was independently studied by
Stückelberg~\cite{stuckelberg1932} and Majorana~\cite{majorana1932}.

\begin{figure}
\includegraphics[width=0.48\textwidth]{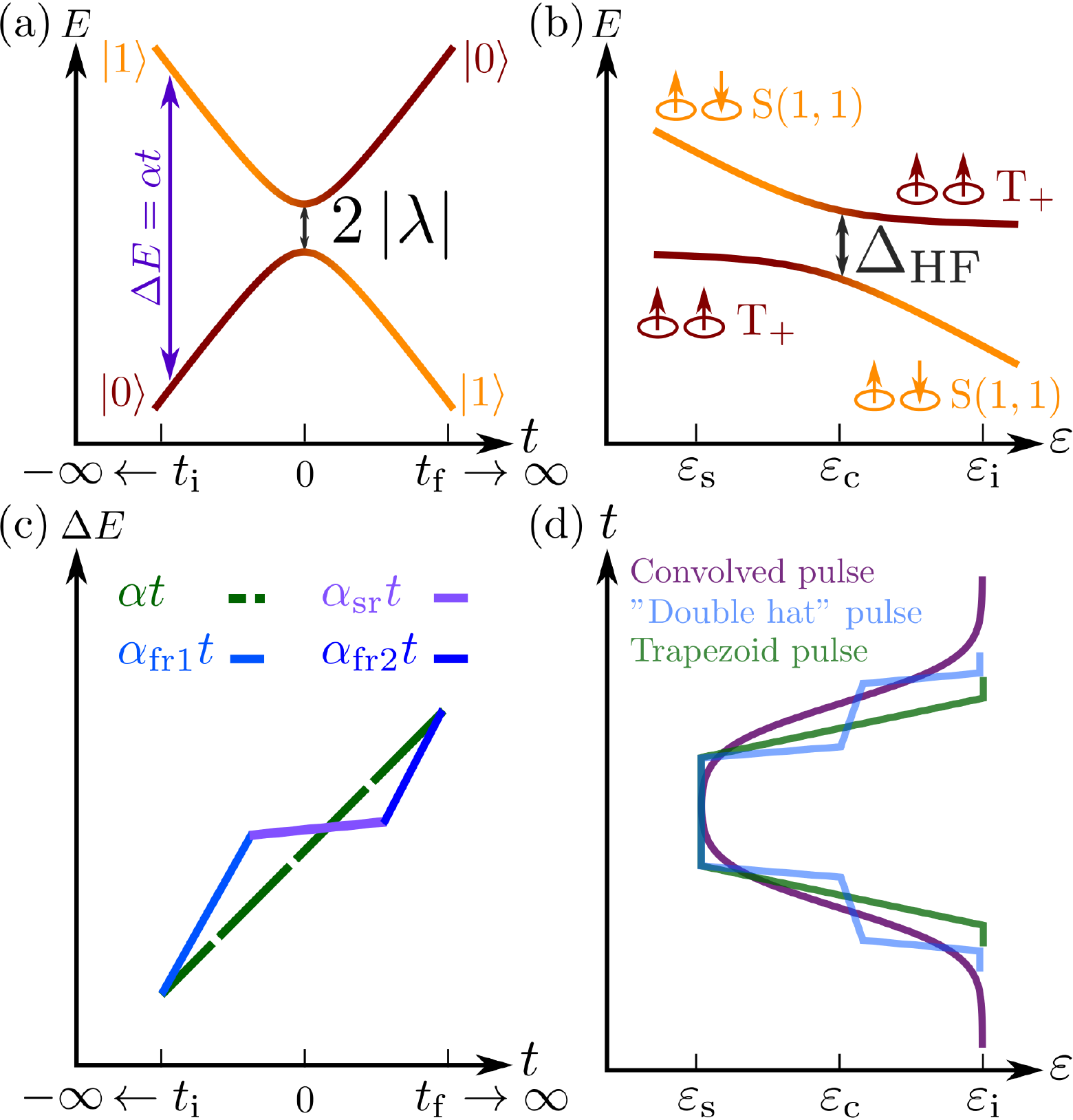}
\caption{(color online) (a) The Landau-Zener-Stückelberg-Majorana (LZSM) problem. The
LZSM model addresses the problem of a two-level system that is swept through an
anti-crossing. It assumes an infinitely long ramp (i.e. propagation from $t_{\mm{i}} \to
-\infty$ to $t_{\mm{f}} \to \infty$), a constant coupling constant $\lambda$ leading to a
splitting $ 2\abs{\lambda}$ at $t=0$, and an energy difference between the
levels that varies linearly with time, $\Delta E(t) = \alpha t$. Its main result gives the
non-adiabatic transition probability, $P_{\mm{LZSM}}$. An extension of the model to
finite-times resolves the problem of infinite energies and undefined phases. (b) Singlet
and triplet $\mm{T}_+$  energies in a DQD as a function of detuning $\varepsilon$, where
$\varepsilon_{\rm i}=\varepsilon(t_{\rm i})$ and $\varepsilon_{\rm s}=\varepsilon(t_{\rm
f})$.  The two-electron DQD is a physical realization of the LZSM model.  Here, the
hyperfine coupling of the two spin states leads to a splitting $\Delta_{\rm HF}$. (c) The
LZSM model assumes a constant level velocity $\alpha$. In order to increase adiabaticity
one can lower $\alpha$, but in order to keep short pulses it is preferable to use multi
rise-time pulses. (d) Comparison of possible pulses that can be used to manipulate
the $\mm{S}$-$\mm{T}_+$ qubit. In terms of total propagation time, ``double hat''
pulses are a good compromise between conventional trapezoid and convolved pulses.}
\label{fig:simplelzsm}
\end{figure}

The Hamiltonian studied in the LZSM model describes a system with two energy levels [see
Fig.~\ref{fig:simplelzsm}(a)] that are coupled by an off-diagonal matrix element
$\lambda$, $H(t)$ = $-(\alpha t/2) \sigma_z$ + $\lambda \sigma_x$, where $\sigma_z$ and
$\sigma_x$ are Pauli matrices, and
$\alpha$ = $\ud(E_2 (t) - E_1 (t)/\ud t$.
The main result of the theory is the asymptotic expression for the non-adiabatic
transition probability when the propagation lasts from $t_{\mm{i}} = -\infty$ to
$t_{\mm{f}} = \infty$,
\begin{equation}
P_{\mm{LZSM}} = \ue^{\frac{- 2 \pi \lambda^2}{\alpha \hbar}}.
\label{eq:simpleLZSM}
\end{equation}

Here we use a generalization of the LZSM model, known as the finite-time LZSM
model~\cite{vitanov1996}. It resolves the problem of the energy divergence when
$t_{\mm{f,i}} = \pm \infty$, and in contrast to the simple case it yields the relative
phase between the states, which is crucial for predicting the coherent time evolution of any
quantum system.  Particularly, knowledge of the relative phase is essential in LZSM
interferometry~\cite{stuckelberg1932,shevchenko2010} in which the system is driven back
and forth across an anti-crossing. The driving generates quantum interference between
states, which is directly observable in the non-adiabatic (or adiabatic) transition
probability.

\subsection{Finite-time Landau-Zener-Stückelberg-Majorana propagator}

The unitary evolution operator defined by the LZSM
Hamiltonian~\cite{landau1932,zener1932,stuckelberg1932,majorana1932},
\begin{equation}
H (t)=
\begin{pmatrix}
-\frac{\alpha}{2} t && \lambda \\[6pt]
\lambda && \frac{\alpha}{2} t
\end{pmatrix},
\label{eq:hlzs}
\end{equation}
is given by~\cite{vitanov1996}
\begin{equation}
U (t_{\mm{f}},\,t_{\mm{i}}) =
\begin{pmatrix}
u_{11} (t_{\mm{f}},\,t_{\mm{i}})  & u_{12}(t_{\mm{f}},\,t_{\mm{i}})\\
u_{21}(t_{\mm{f}},\,t_{\mm{i}}) & u_{22}(t_{\mm{f}},\,t_{\mm{i}})
\end{pmatrix},
\label{eq:lzsm_matrix}
\end{equation}
where $1 (2)$ refers to the states $\ket{0}$ ($\ket{1}$) with
\begin{equation}
\begin{split}
&u_{11} (t_{\mm{f}},\,t_{\mm{i}}) = u_{22}^{\ast}(t_{\mm{f}},\,t_{\mm{i}})=\\
& \frac{\Gamma\left(1 - \ui\eta^2\right)}{\sqrt{2\pi}}
\left[D_{\ui \eta^2}\left(\ue^{\frac{-\ui\pi}{4}}\tau_{\mm{f}}\right)
D_{\ui\eta^2-1}\left(\ue^{\frac{3\ui\pi}{4}}\tau_{\mm{i}}\right)
\right.\\
&+ \left. D_{\ui \eta^2}\left(
\ue^{\frac{3\ui\pi}{4}}\tau_{\mm{f}}\right)
D_{\ui\eta^2-1}\left(\ue^{-\frac{\ui\pi}{4}}\tau_{\mm{i}}\right)
\right],
\end{split}
\label{eq:u11}
\end{equation}
and
\begin{equation}
\begin{split}
&u_{12} (t_{\mm{f}},\,t_{\mm{i}}) = -u_{21}^{\ast}(t_{\mm{f}},\,t_{\mm{i}})=\\
& \frac{\Gamma\left(1 - \ui\eta^2\right)}{\sqrt{2 \pi}\eta}
\ue^{\frac{\ui\pi}{4}}
\left[-D_{\ui \eta^2}\left(\ue^{\frac{-\ui\pi}{4}}\tau_{\mm{f}}\right)
D_{\ui\eta^2}\left(\ue^{\frac{3\ui\pi}{4}}\tau_{\mm{i}}\right)
\right.\\
&+ \left. D_{\ui \eta^2}\left(
\ue^{\frac{3\ui\pi}{4}}\tau_{\mm{f}}\right)
D_{\ui\eta^2}\left(\ue^{-\frac{\ui\pi}{4}}\tau_{\mm{i}}\right)
\right].
\end{split}
\label{eq:u12}
\end{equation}
Here $\tau = \sqrt{\alpha/\hbar}t$ is a dimensionless time, $\eta = \lambda/\sqrt{\alpha
\hbar}$ is a dimensionless coupling, $\Gamma (z)$ is the gamma
function~\cite{abramowitz_gamma}, and $D_{\nu} (z)$ is the parabolic cylinder
function~\cite{erdelyi_parabolic}. By definition, $t=0$ is set at the anti-crossing. The
usual LZSM formula eq.~\eqref{eq:simpleLZSM} is retrieved from the modulus square of
Eq.~\eqref{eq:u11} when taking the limit $t_{\mm{i}} \to -\infty$ and $t_{\mm{f}} \to
\infty$.

The LZSM propagator fully determines the partial adiabatic dynamics of a quantum
two-level system. In the case where the two-level system encodes a qubit,
Eq.~\eqref{eq:lzsm_matrix} allows the design of single qubit operations. This method has
been used to control superconducting qubits~\cite{oliver2005,shevchenko2010}, and more
recently to implement a qubit encoded in the spin of a two-electron
state~\cite{petta2010,ribeiro2010}. However, since the spin states are weakly coupled,
$\eta < 1$, it is hard to achieve an equally weighted coherent superposition of spin
states and fully explore the entire qubit state space. In order to achieve full control
over the spin qubit, it would be necessary to perform slower sweeps, i.e., to increase
$\eta$ by making $\alpha$ smaller.  However, the LZSM equation requires an exponential
increase in the propagation time in order to achieve a fully adiabatic transition. In a
real physical system, this method is unpractical because the pulse duration needs to
remain shorter than the coherence time of the two-level system.

\subsection{Observing finite-time effects}

As demonstrated in Ref.~\onlinecite{ribeiro2012}, it is possible to use more complex
pulses to increase the balance of the populations while keeping the manipulation time
below the decoherence times. The key idea relies on an observation based on the
finite-time LZSM model. For a slow-level velocity $\alpha$, which favors adiabatic
passage, most of the population change occurs in the vicinity of the anti-crossing. It is
therefore possible to use detuning pulses that have a time-dependent level velocity. Let
us consider two types of pulses, as illustrated in Figs.~\ref{fig:simplelzsm}(c) and (d).
The first is a conventional linear pulse, which is standard in LZSM theory. The second
pulse profile consists of linear detuning ramps in a fast-slow-fast rise-time sequence,
which we refer to as ``double hat'' pulse.  The unitary evolution of such a general
sequence can be written using Eq.~\eqref{eq:lzsm_matrix} as
\begin{equation}
\begin{aligned}
U(t_{\mm{f}}, t_{\mm{i}}) &= U_{\mm{fast2}}(t_{\mm{f}},\, t_2)
U_{\mm{slow}}(t_2,\, t_1)
U_{\mm{fast1}}(t_1,\, t_{\mm{i}})\\
&=
\begin{pmatrix}
\tilde{u}_{11} (t_{\mm{f}},\,t_{\mm{i}})  & \tilde{u}_{12}(t_{\mm{f}},\,t_{\mm{i}})\\
\tilde{u}_{21}(t_{\mm{f}},\,t_{\mm{i}}) & \tilde{u}_{22}(t_{\mm{f}},\,t_{\mm{i}})
\end{pmatrix}.
\end{aligned}
\label{eq:doublehatprop}
\end{equation}
Here $t_1 = t_{\mm{i}} + t_{\mm{fr1}}$, $t_2 = t_{\mm{i}} + t_{\mm{fr1}} + t_{\mm{sr}}$,
and $t_{\mm{f}} = t_{\mm{i}} + t_{\mm{fr1}} + t_{\mm{sr}} + t_{\mm{fr2}}$, where
$t_{\mm{fr}j}$ is the propagation time associated to the $j$th fast sequence, and
$t_{\mm{sr}}$ corresponds to the slow sequence. We use this notation to refer to the
corresponding level velocities $\alpha_j$, dimensionless times $\tau_j$, and
dimensionless couplings $\eta_j$.

In addition to the already mentioned and studied advantages ``double hat'' pulses offer,
they also provide sensitive means to explore finite-time effects. These are in general
neglected when describing experiments because the more convenient LZSM scattering
approach~\cite{shevchenko2010} has been sufficient to reproduce experimental
results~\cite{oliver2005,petta2010}.  However, to implement high-fidelity quantum gates
it will be necessary to accurately describe the dynamics of the qubit and thus take into
account finite-time propagation.

\begin{figure}
\includegraphics[width=0.48\textwidth]{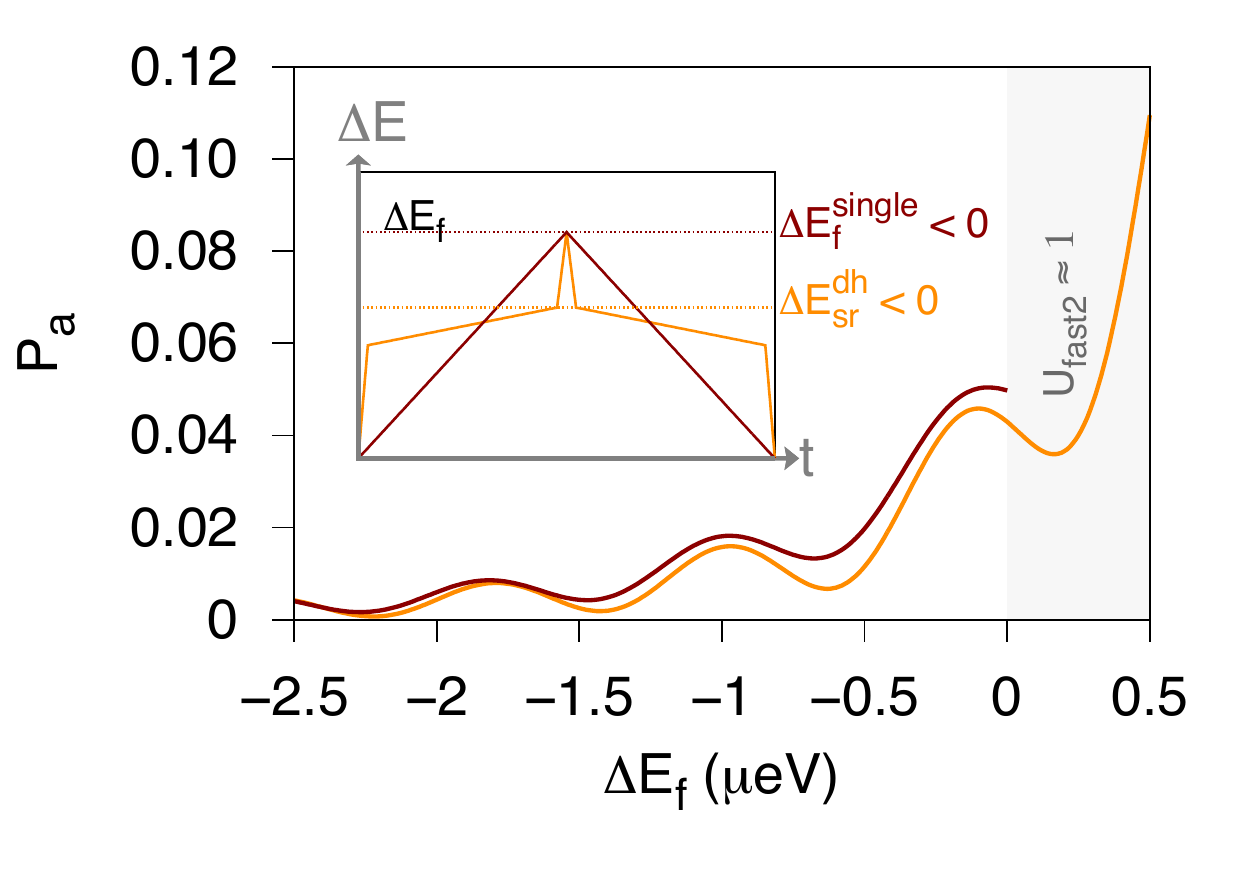}
\caption{(color online) Comparison of the adiabatic transition probability, $P_{\mm{a}}$,
for a ``double hat'' (orange) and a linear pulse (red) with $t_{\mm{w}}=0$ as a function
of $\Delta E_{\mm{f}}$. (Inset) Pulse profiles used to obtain $P_{\mm{a}}$. The maximal
offset for the single rise-time pulse is chosen such that the system is never driven
through the anti-crossing ($\Delta E_{\mm{f}}^{\mm{single}} <0$).
The maximal offset for ``double hat'' pulses is determined by the condition that the
system cannot be driven with the slow component of the pulse through the anti-crossing
($\Delta E_{\mm{sr}}^{\mm{dh}} <0$). For ``double hat'' pulses, the values of
$P_{\mm{a}}$ between $0 < \Delta E_{\mm{f}} < 0.5\,\mm{\mu eV}$ originate from the slow
portion of the pulse, which brings the system close to the anti-crossing. In this range,
the magnitude of $P_{\mm{a}}$ is not due to the second fast rise-time portion of the
pulse, which drives the system through the anti-crossing.}
\label{fig:dhlzsm}
\end{figure}

In Fig.~\ref{fig:dhlzsm} we compare adiabatic transition probabilities obtained with
``double hat'' pulses and conventional trapezoid pulses [c.f. inset of
Fig.~\ref{fig:dhlzsm}]. Here, an adiabatic transition refers to a transition where the
system remains in an instantaneous energy eigenstate. The details of the leading
edge (the trailing edge is identical but reversed) of the ``double hat'' pulse are $t_{\mm{fr1}} =
t_{\mm{fr2}} = 0.1\,\mm{ns}$, the starting position of the slow-level velocity ramp is defined
by the condition $\Delta E_{\mm{i}} - \Delta E_{\mm{i,sr}} = -2\,\mm{\mu eV}$, where
$\Delta E_{\mm{i}}$ is the initial energy difference between the uncoupled eigenstates
and $\Delta E_{\mm{i,sr}}$ the energy difference at the beginning of the slow part of the
pulse. We choose $\alpha_{\mm{sr}} = 500\,\mm{eVs^{-1}}$ and $\lambda = 70\,\mm{neV}$. We
impose the same total propagation time on the linear pulse as for the ``double hat'',
from which we obtain the level velocity $\alpha_{\mm{si}}$ for the linear pulse,
\begin{equation}
	\alpha_{\mm{si}} = \frac{\Delta E_{\mm{f}} - \Delta
	E_{\mm{i}}}{t_{\mm{fr1}} + t_{\mm{fr2}} + t_{\mm{sr}}}.
	\label{eq:alphasimple}
\end{equation}
The adiabatic transition probability, $P_{\mm{a}}$, is plotted as a function of
$\Delta E_{\mm{f}}$. Here, $\Delta E_{\mm{f}}$ indicates the energy difference between
the states when the pulse has reached its maximal amplitude. Different values of $\Delta
E_{\mm{f}}$ are obtained by adding an offset to $\Delta E_{\mm{i}}$, while keeping the
length of pulse (thus $\alpha_{\mm{si}}$ and the different $\alpha_j$ of the ``double
hat'') constant. Since we are interested in finite-time effects, we impose $\Delta
E_{\mm{f}} < 0$ for trapezoid pulses, which reflects that the system is not detuned
through the anti-crossing. This condition is relaxed for ``double hat'' pulses, for which
we impose $\Delta E_{\mm{f}} < 0.5\,\mm{\mu eV}$. This is equivalent, with our choice of
parameters, to letting the system be driven up to the anti-crossing with the slow
component of the ``double hat''.  This condition can be written as $\Delta
E_{\mm{sr}}^{\mm{dh}} < 0$, i.e. the energy difference at the end of the slow rise-time
component is smaller than $0$.

Our results show for the trapezoid pulse what is expected from a finite-time LZSM
theory~\cite{vitanov1996}. There is a small probability for an adiabatic transition if
the system is detuned to close proximity of the anti-crossing. If one compares this
result with values of $P_{\mm{a}}$ obtained with the ``double hat'', it seems that there
is no enhancement. Moreover, one would have a tendency to associate values of
$P_{\mm{a}}$ between $0 < \Delta E_{\mm{f}} < 0.5\,\mm{\mu eV}$ as originating from the
second fast-rise portion of the pulse, which drives the system through the anti-crossing.
However, for the particular ``double hat'' we are considering here, we have
$U_{\mm{fast2}} \approx \mathbbm{1}$. This means that the values of $P_{\mm{a}}$ in the
range $0 < \Delta E_{\mm{f}} < 0.5\,\mm{\mu eV}$ are due to the system being brought
close to the anti-crossing with the slow portion of the pulse, for which $\eta \gtrsim
1$. This is in contrast with trapezoid pulses, or any single rise-time pulse, for which
$\Delta E_{\mm{f}}>0$ implies that most of the magnitude of $P_{\mm{a}}$ comes from the
system being driven through the anti-crossing.

\begin{figure}
\includegraphics[width=0.48\textwidth]{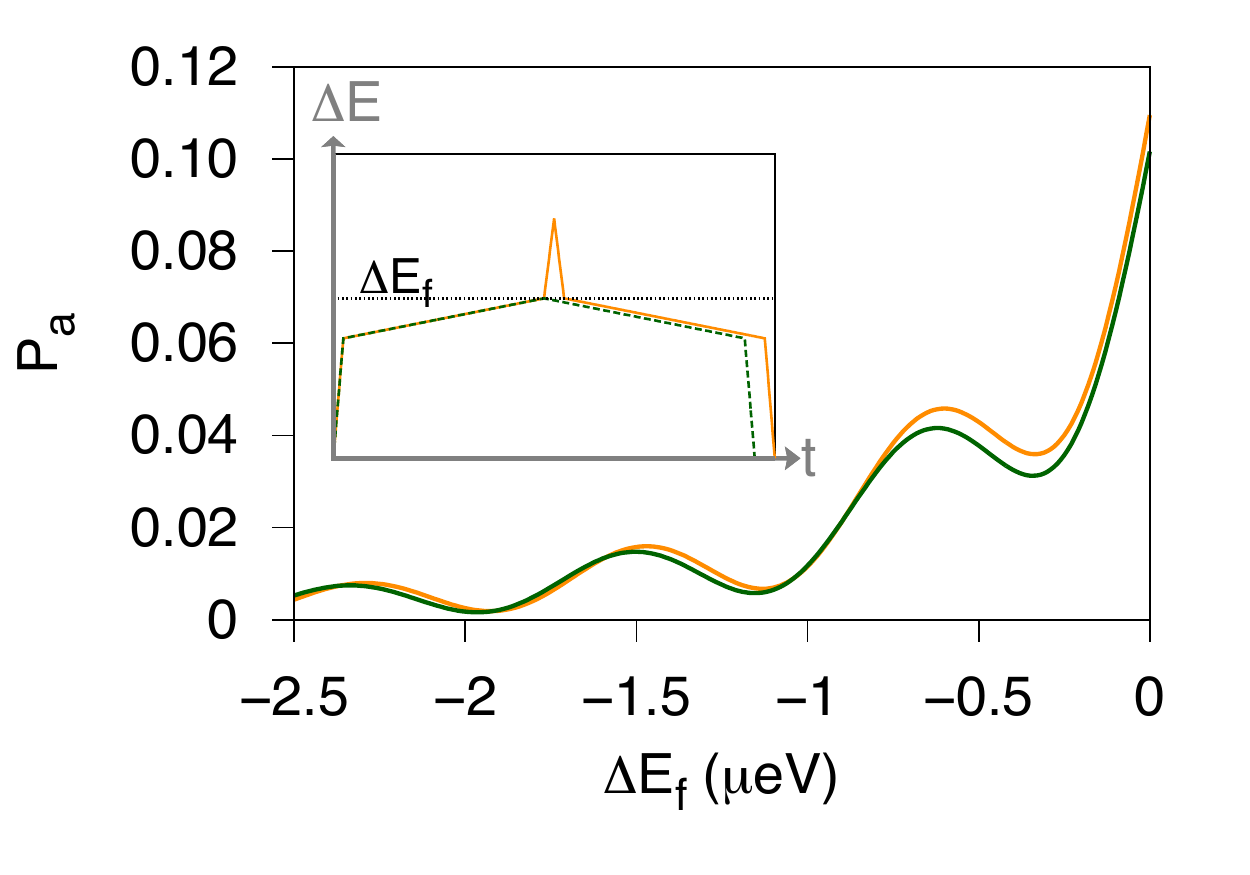}
\caption{(color online) Comparison of the adiabatic transition probability,
$P_{\mm{a}}$, for a ``double hat'' (orange) and a ``truncated double hat''
(green), which is missing the second fast rise-time component, as a function of
$\Delta E_{\mm{f}}$ at the end of the slow detuning pulse. (Inset) Pulse profiles
used to obtain $P_{\mm{a}}$. The results obtained with a ``double hat'' are
shifted by an amount $\Delta E_{\mm{fast}2}$ on the $x$-axis to allow for
comparison. We conclude from these results that the state of the system is hardly
changed during the second fast detuning pulse.}
\label{fig:comp_finite-time}
\end{figure}

To illustrate our last statement, in Fig.~\ref{fig:comp_finite-time} we present a
comparison between $P_{\mm{a}}$ obtained with a ``double hat'' pulse as described
previously and a ``truncated double hat'', which is missing the second fast detuning
ramp [c.f. inset Fig.~\ref{fig:comp_finite-time}]. To compare $P_{\mm{a}}$ between the two
pulses, we choose the $x$-axis to describe $\Delta E_{\mm{f}}$ at the end of the slow
detuning pulse. We shift the values obtained with a ``double hat'' along the $x$-axis by
an amount $\Delta E_{\mm{fast}2} = 0.5\,\mm{\mu eV}$ to compare between the two different
pulses.  We clearly see that for both cases $P_{\mm{a}}$ is nearly identical.

\section{Adiabatic Control of a $\mm{S}-\mm{T}_+$ qubit}
\label{sec:st+}

In the following, we apply the previously developed ideas to a physical implementation of
a LZSM driven qubit.  We focus on the two-spin $\mm{S}-\mm{T}_+$ implementation in a GaAs
DQD~\cite{petta2010,ribeiro2010}, see Fig.~\ref{fig:simplelzsm}(b). Although the
dynamics of the system under ``double hat'' pulses has already been studied
experimentally~\cite{ribeiro2012}, there is still a need to gain a better understanding
of the charge-noise-induced spin dephasing. We will show, among other things, that the
measurement of finite-time LZSM oscillations can provide a tool to qualitatively access
the strength of charge noise.

\subsection{Double Quantum Dot Spin States}

The spin preserving part of the Hamiltonian describing the confinement of electrons in a
DQD in the presence of an external magnetic field can be written using a simple two-site
hopping model,
\begin{equation}
\begin{aligned}
H_0 &= \sum_{\substack{{i=1,2}\\{\sigma=\uparrow,\,\downarrow}}}\left(\varepsilon_i +
\frac{1}{2} g^* \mu_{\mm{B}}B \sigma\right)c_{i \sigma}^{\dag} c_{i \sigma} + u \sum_i c_{i \uparrow}^{\dag}
c_{i \uparrow} c_{i \downarrow}^{\dag} c_{i \downarrow}\\
&\phantom{=} + \tau \sum_{\sigma} \left(c_{1 \sigma}^{\dag} c_{2 \sigma} + \mm{h. c.}\right).
\end{aligned}
\label{eq:DQD_H0}
\end{equation}
The index $i=1,\,2$ labels the dot number and $\sigma=\uparrow,\downarrow=\pm 1$ the spin
of the electron. We denote the energy of a single confined electron by $\varepsilon_i$
and the Zeeman energy associated with its spin is given by $g^* \mu_{\mm{B}}B \sigma/2$,
where $g^{*}$ denotes the effective Landé g-factor, $\mu_{\mm{B}}$ the Bohr magneton, $B$
the strength of the external magnetic field. The operators $c_{i \sigma}$ and $c_{i
\sigma}^{\dag}$ describe respectively the annihilation and creation of an electron in dot
$i$ with spin $\sigma$. Two electrons occupying the same QD give rise to an intra-dot
Coulomb energy $u$.  The last term of Eq.~\eqref{eq:DQD_H0} accounts for electron
tunneling between the dots with strength $\tau$. We neglect the inter-dot Coulomb
interaction because it only produces a constant shift of the energy levels.

Since most recent experiments on a  DQD system are operated in a regime with at most two
electrons, we can project Eq.~\eqref{eq:DQD_H0} into the subspace spanned by the charge
configurations $(0,2)$, $(2,0)$, and
$(1,1)$~\cite{petta2005,foletti2009,petta2010,bluhm2011,studenikin2012}. The diagonalization of the
resulting Hamiltonian leads to six low-energy states which are superpositions of the
singlets $\mm{S}(0,2)$, $\mm{S}(2,0)$, and $\mm{S}(1,1)$ as well as the triplets
$\mm{T}_0 (1,1)$, $\mm{T}_+ (1,1)$, and $\mm{T}_- (1,1)$. The triplet states with two
particles in the same dot must have electrons occupying higher-energy orbitals due
to the Pauli principle. This results in a relatively high separation in energy ($\sim
400\,\mu\mm{eV}$) from a singlet with both electrons occupying the same
dot~\cite{hanson2004}. Consequently the triplets states with two electrons in the same
dot can be safely neglected for the purpose of the current study.

\begin{figure}
\includegraphics[width=0.48\textwidth]{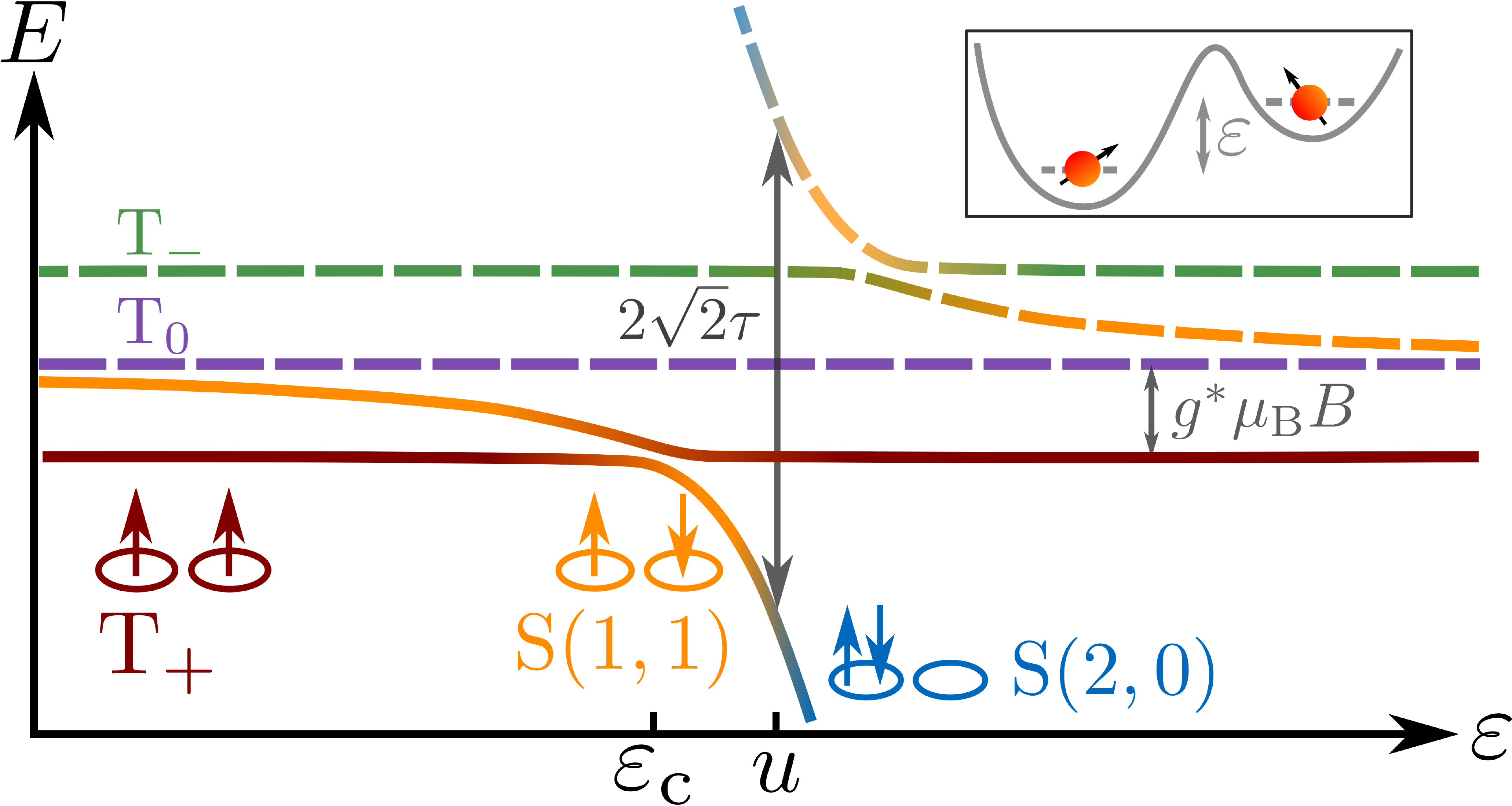}
\caption{(color online) Energy diagram for the relevant states in the DQD as a
function of $\varepsilon$. The spin states for the implementation of the qubit are the
hybridized singlet $\mm{S}$ and the triplet $\mm{T}_+$.}
\label{fig:energy_diagram}
\end{figure}

An energy level diagram is shown in Fig.~\ref{fig:energy_diagram}, where we plot the
energies of the relevant two-electron spin states as a function of detuning $\varepsilon
= \varepsilon_1 - \varepsilon_2$.
The degeneracy of the singlets $\mm{S}(1,1)$ and $\mm{S}(2,0)$ at $\varepsilon =u$, as
well as between $\mm{S}(1,1)$ and $\mm{S}(0,2)$ at $\varepsilon=-u$, is lifted due to
tunneling, which results in a splitting of the levels by $2 \sqrt{2} \tau$.  The
degeneracy between the spin singlet $\mm{S}$ and spin triplet $\mm{T}_0$ is lifted due to
the exchange interaction~\cite{burkard1999}. This property has allowed the encoding of a
spin qubit in these two spin states and its manipulation via the exchange
interaction~\cite{petta2005}.

Here we concentrate on a particular value of detuning that we have denoted by
$\varepsilon_{\mm{c}}$ in the energy level diagram. This point corresponds to the
crossing of the singlet state with the triplet $\mm{T}_+$ state and is special because
there is an anticrossing due to the hyperfine interaction between the electron spins and
nuclear spins of the host material. It has been demonstrated
experimentally~\cite{petta2010} and theoretically~\cite{ribeiro2010} that coherent
control of the $\mm{S}$-$\mm{T}_+$ qubit can be achieved by detuning the system from an
initially prepared $\mm{S} (2,0)$ through the hyperfine mediated anti-crossing.

The hyperfine interaction is described by the effective Hamiltonian
\begin{equation}
H_{\mm{HF}} = \Sv_1 \cdot \hv_1 + \Sv_2 \cdot \hv_2
\label{eq:hfgen}
\end{equation}
between the electron spin $\Sv_i$ and the effective magnetic fields $\hv_i$ that are
generated by the nuclear spins $\Iv_i$ in dot $i$. The Overhauser field operators $\hv_i
= \sum_{k=1}^{n_i} A_i^k \Iv_i^k$ describe the nuclear spin bath. Here $n_i$ is the number of
nuclei in dot $i$ and $A_i^k = v_{i k} \nu_0 \left|\psi_i (\rv_k)\right|^2$ is the
hyperfine coupling constant with the $k$-th nucleus in dot $i$, with $\psi_i (\rv_k)$ the
electron wave function, $\nu_0$ the volume of the unit cell and $v_{ik}$ the hyperfine
coupling strength. A more convenient form of the hyperfine interaction is obtained by
introducing the spin ladder operators $S_i^\pm=S_i^x\pm \mm{i} S_i^y$ and $h_i^{\pm} =
h_i^x \pm \ui h_i^y$, which yields
\begin{equation}
H_{\mm{HF}} = \frac{1}{2} \sum_i \left(2 S_i^z
h_i^z + S_i^+ h_i^- + S_i^- h_i^+\right).
\label{eq:hfladder}
\end{equation}
The longitudinal part of $H_{\mm{HF}}$ is diagonal and its contribution will add to the
energy of the triplet state. The transverse part
\begin{equation}
	H_{\mm{HF}}^{\perp} = \frac{1}{2}\sum_i (S_i^+ h_i^- + S_i^- h_i^+),
\label{eq:hftrans}
\end{equation}
generates the so-called flip-flop process that result in an energy gap at
$\varepsilon_{\mm{c}}$ and allows for mixing of the $\mm{S}$ and $\mm{T}_+$ spin
states~\cite{khaetskii2002,coish2005,taylor2007}.

\subsection{$\mm{S}-\mm{T}_+$ Effective Hamiltonian}
\label{sec:chargespin}

In this section, we derive an effective $2\times2$ Hamiltonian that describes the
dynamics of the $\mm{S}-\mm{T}_+$ spin states near the hyperfine induced anti-crossing.
Before doing so, we start by making a few considerations based on LZSM theory to
determine which states play a negligible role in the dynamics. Since Eq.~\eqref{eq:DQD_H0} describes a series of
anti-crossings, it is possible to use the results of Refs.~\onlinecite{kayanuma1985,usuki1997} where a
formula for the non-adiabatic transition probability of a multiple crossings LZSM model has
been derived. It was shown that for well separated anti-crossings~\cite{usuki1997}, we
have
%
	$P_k = \prod_{j=1}^k \exp(- 2 \pi \lambda_j^2  /  \hbar \alpha_j)$,
%
where $P_k$ is the non-adiabatic transition probability after the $k$-th anti-crossing. This
formula is a product of $k$ LZSM probabilities, which reflects the independence between
the set of anti-crossings.

This model can directly be applied to the DQD system for magnetic fields on the order of
a few hundreds of mT, where the charge anti-crossing and both of the hyperfine induced
anti-crossings ($\mm{T}_+$ and $\mm{T}_-$) are well separated. Here, we demonstrate that
if the system is initialized in the singlet $\mm{S}(2,0)$, then the detuning pulses
allowing for mixing of the lowest energy hybridized singlet state and triplet $\mm{T}_+$
cannot populate the higher energy hybridized singlet state and consequently the triplet
$\mm{T}_-$. Let us consider that the hyperfine coupling strength is on the order of a
hundred nano-electron volts, $\lambda_{\mm{HF}} = 100\,\mm{neV}$. This value is
consistent with experimental findings. In Ref.~\onlinecite{petta2010}, a strength of
$60\,\mm{neV}$ has been reported. For our choice of $\lambda_{\mm{HF}}$, we find
$P_{\mm{LZSM}} = 0.5$ for $\alpha \simeq 138\,\mm{eVs^{-1}}$. This would correspond to an
equally weighted superposition of the qubit states. However, this is only true if there
is no population transfer to the higher hybridized singlet level as predicted by
the equation for $P_k$. By evaluating the population transfer between $\mm{S}(2,0)$ and
$\mm{S}(1,1)$ with the LZSM formula for $\alpha = 138\,\mm{eVs^{-1}}$ and
$\lambda_{\mm{charge}} = \sqrt{2} \tau \simeq 7.1\,\mm{\mu eV}$, we find $P_{\mm{LZSM}}
\simeq 0$, within the numerical precision of our calculation, which indicates that there
is no population transfer.

From the previous considerations, we have shown that it is safe to neglect the higher
energy hybridized singlet state as well as the triplet $\mm{T}_-$, but the importance of
$\mm{T}_0$ remains to be determined. Here, we rely on recent experimental
results~\cite{studenikin2012} which demonstrate that only a certain type of detuning
pulses lead to mixing between $\mm{T}_+$, $\mm{T}_0$, and the ground-state
singlet. Moreover, as the results of Ref.~\onlinecite{studenikin2012} indicate, it is
possible to identify in interference patterns the presence of $\mm{T}_0$ in the
dynamics~\cite{sarkka2011}.

Finally, we conclude that it is possible to restrict the Hilbert space
to two states, $\mm{T}_+ (1,1)$ and the lowest energy hybridized
singlet S.
In order to derive
an analytical expression for the latter, we start by considering the projection of
Eq.~\eqref{eq:DQD_H0} onto the states $\mm{T}_+ (1,1)$, $\mm{S}(1,1)$, and $\mm{S}(2,0)$.
We find
\begin{equation}
H_0 (\varepsilon) =
\begin{pmatrix}
g^* \mu_{\mm{B}} B & 0 & 0\\
0 & 0 & \sqrt{2} \tau\\
0 & \sqrt{2} \tau & u - \varepsilon \\
\end{pmatrix}.
\label{eq:3x3ham}
\end{equation}

The diagonalization of Eq.~\eqref{eq:3x3ham} yields two hybridized singlets and the
triplet state $\mm{T}_+$
\begin{empheq}{align}
&\ket{\mm{S}} = c(\varepsilon) \ket{\mm{S}(1,1)} + \sqrt{1-c(\varepsilon)^2}\ket{\mm{S}(2,0)},
\label{eq:gsinglet}\\
&\ket{\mm{S'}} = c'(\varepsilon) \ket{\mm{S}(1,1)} +
\sqrt{1- c'(\varepsilon)^2}\ket{\mm{S}(2,0)},
\label{eq:esinglet}\\
&\ket{\mm{T}} = \ket{\mm{T}_+ (1,1)},
\label{eq:triplet}
\end{empheq}
with respective energies
\begin{empheq}{align}
&E_{\mm{S}} (\varepsilon)  = \frac{1}{2}(u -
\varepsilon - \sqrt{8 \tau^2 + (u - \varepsilon)^2}),
\label{eq:energ_gsinglet}\\
&E_{\mm{S'}} (\varepsilon) = \frac{1}{2}(u - \varepsilon + \sqrt{8 \tau^2 + (u -
\varepsilon)^2}),
\label{eq:energ_esinglet}\\
&E_{\mm{T}} = g^* \mu_{\mm{B}} B.
\end{empheq}
The charge admixture coefficients $c(\varepsilon)$ and
$c'(\varepsilon)$ are
\begin{empheq}{align}
c(\varepsilon) &= \frac{\varepsilon - u -
\sqrt{8 \tau^2 + (u - \varepsilon)^2}}{\sqrt{8 \tau^2 + (-\varepsilon + u + \sqrt{8
\tau^2 + (\varepsilon -u)^2})^2}},
\label{eq:cepsilon}\\
c'(\varepsilon) &= \frac{\varepsilon - u + \sqrt{8
\tau^2 + (u - \varepsilon)^2}}{\sqrt{8 \tau^2 + (\varepsilon - u + \sqrt{8 \tau^2 +
(\varepsilon -u)^2})^2}}.
\label{eq:cepsilonprime}
\end{empheq}

We can now make the following basis change
\begin{equation}
T(\varepsilon) =
\begin{pmatrix}
1 & 0 & 0\\
0 & c(\varepsilon) & c'(\varepsilon)\\
0 & \sqrt{1 - c^2(\varepsilon)}& \sqrt{1 - c'^2(\varepsilon)} \\
\end{pmatrix},
\label{eq:matchangbasis}
\end{equation}
such that
\begin{equation}
\begin{aligned}
H_0^{\mm{diag}} (\varepsilon) &=
T(\varepsilon)^{\dag} H_0 (\varepsilon) T(\varepsilon)\\
&=
\begin{pmatrix}
g^* \mu_{\mm{B}} B & 0 & 0\\
0 & E_{\mm{S}} & 0\\
0 & 0 & E_{\mm{S'}}\\
\end{pmatrix}.
\end{aligned}
\label{eq:diag3x3ham}
\end{equation}

We add now to the Hamiltonian defined in Eq.~\eqref{eq:3x3ham} the hyperfine interaction
defined in Eq.~\eqref{eq:hfgen}.  Since we are not interested in describing the nuclear
spins dynamics, but its effect on the two-spin states $\mm{S}$ and $\mm{T}_+$, we can
model the action of the Overhauser field operators by introducing a classical stochastic
variable which accounts for fluctuations in the nuclear
spin ensemble~\cite{khaetskii2002,coish2005,taylor2007}. By setting
\begin{equation}
\hv_i = g^* \mu_{\mm{B}} \mathbf{B}_{\mm{n},i},
\label{eq:semiclOver}
\end{equation}
we can interpret $\mathbf{B}_{\mm{n},i}$ as the effective random magnetic field
acting on $\Sv_i$.  Under normal experimental conditions we have $k_{\mm{B}} T \gg
g_{\mm{n}} \mu_{\mm{n}} B$, where $g_{\mm{n}}$ and $\mu_{\mm{n}}$ are the nuclear
$g$-factor and magneton. The nuclear spins can, in this limit, be assumed to be completely
unpolarized, resulting in a Gaussian distribution of nuclear
fields~\cite{khaetskii2002,coish2005,taylor2007}
\begin{equation}
	p(B_{\mm{n},\,i}) = \frac{1}{\sqrt{2 \pi}
	\sigma}\mm{e}^{-\frac{B_{\mm{n},\,i}^2}{ 2\delta^2}},
\label{eq:nuclfielddist}
\end{equation}
with $\delta = A / g^* \mu_{\mm{B}} \sqrt{n}$, the hyperfine coupling constant $A
\approx 90\,\mm{\mu eV}$, and the approximate number of nuclei overlapping with the
electronic wave function $n \approx 10^5 - 10^6$.  By defining $B_{\mm{n},\,i}^{\pm} =
B_{\mm{n},\,i}^x \pm \mm{i} B_{\mm{n},\,i}^y$, the hyperfine Hamiltonian can be written
by analogy with Eq.~\eqref{eq:hfladder} as
\begin{equation}
H_{\mm{HF}} = \frac{1}{2} \sum_i \left(2 S_i^z
B_{\mm{n},i}^z + S_i^+ B_{\mm{n},i}^- + S_i^- B_{\mm{n},i}^+\right),
\label{eq:hfladdersc}
\end{equation}
\begin{figure}
\includegraphics[width=0.48\textwidth]{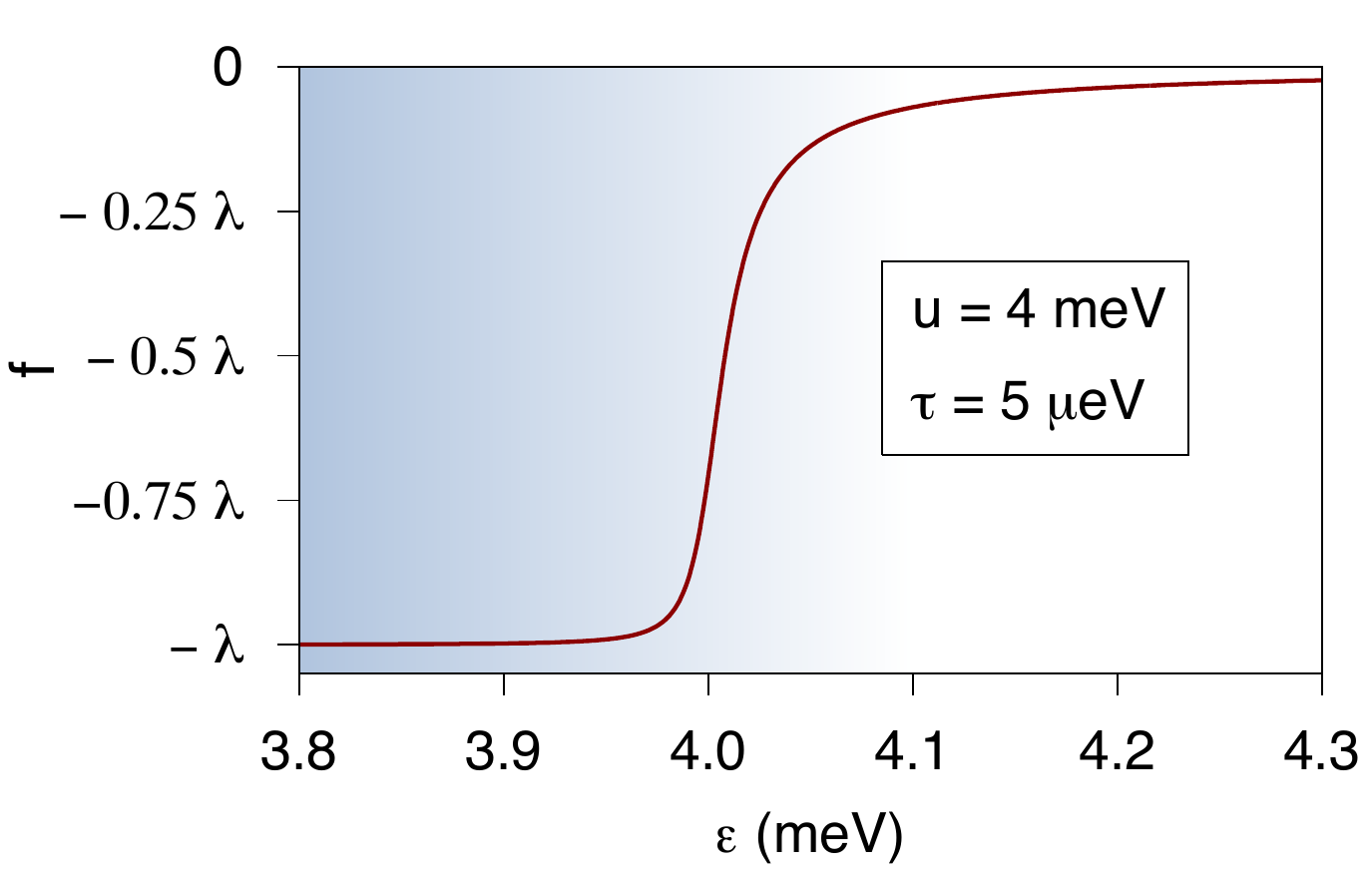}
\caption{(color online) The effective hyperfine mediated coupling between $\ket{\mm{S}}$
and $\ket{\mm{T}}$ is a function of detuning. This is a consequence of
the state $\ket{\mm{S}}$ being a superposition of different charge states.}
\label{fig:coupling}
\end{figure}
The longitudinal part of $H_{\mm{HF}}$ can be included in the energy of the triplet state
$g^* \mu_{\mm{B}} B \to g^* \mu_{\mm{B}} (B + B_{\mm{n},1}^z +  B_{\mm{n},2}^z)$, while the
transverse part
\begin{equation}
	H_{\mm{HF}}^{\perp} = \frac{1}{2}\sum_i (S_i^+ B_{\mm{n},i}^- + S_i^-
B_{\mm{n},i}^+).
\label{eq:hftranscl}
\end{equation}
mixes the spin states $\ket{\mm{S}}$ and $\ket{\mm{T}}$.
We therefore find that the full Hamiltonian, in the $\{\mm{T}_+ (1,1), \mm{S}(1,1),
\mm{S}(2,0)\}$ basis, describing the dynamics of the singlet $\mm{S}$
and triplet $\mm{T}_+$ near the hyperfine anti-crossing is given by
\begin{equation}
H_{\mm{S}-\mm{T}_+} (\varepsilon)  =
\begin{pmatrix}
g^* \mu_{\mm{B}} \tilde{B} & \lambda & 0 \\
\lambda & 0 & \sqrt{2}\tau \\
0 & \sqrt{2} \tau & (u-\varepsilon) \\
\end{pmatrix},
\label{eq:fullH}
\end{equation}
where $\tilde{B} = B + B_{\mm{n},1}^z +  B_{\mm{n},2}^z$ and
\begin{equation}
\lambda = \bra{\mm{S}(1,1)} H_{\mm{HF}}^{\perp} \ket{\mm{T}_+ (1,1)} = g^* \mu_{\mm{B}} (B_{\mm{n},2}^- -
B_{\mm{n},1}^-)/2 \sqrt{2}.
\label{eq:effective_hf_coupling}
\end{equation}
In the basis that diagonalizes Eq.~\eqref{eq:3x3ham}, the Hamiltonian defined in
Eq.~\eqref{eq:fullH} reads
\begin{equation}
\tilde{H}_{\mm{S}-\mm{T}_+} (\varepsilon)  =
\begin{pmatrix}
g^* \mu_{\mm{B}} \tilde{B} & c(\varepsilon) \lambda & c'(\varepsilon) \lambda\\
c(\varepsilon) \lambda & E_{\mm{S}} (\varepsilon) & 0\\
c'(\varepsilon) \lambda & 0 & E_{\mm{S'}} (\varepsilon)\\
\end{pmatrix}.
\label{eq:fullHnew}
\end{equation}

The projection of Eq.~\eqref{eq:fullHnew} onto the Hilbert space spanned by
$\{\ket{\mm{S}},\ket{\mm{T}}\}$ leads to the effective $2 \times 2$ Hamiltonian
describing the dynamics at the singlet-triplet $\mm{T}_+$ anti-crossing.
Taking into account that the detuning is time-dependent, $\varepsilon = \varepsilon (t)$,
we finally obtain
\begin{equation}
H (t)=E_{\mm{S}}(\varepsilon(t))\ket{\mm{S}}\bra{\mm{S}} +
\tilde{E}_{\mm{T}}\ket{\mm{T}}\bra{\mm{T}} +
f(\varepsilon(t))\left(\ket{\mm{S}}\bra{\mm{T}} +
\mm{h. c.}\right),
\label{eq:eff_ham}
\end{equation}
where $E_{\mm{S}}$ is defined in Eq.~(\ref{eq:energ_gsinglet}) and
$\tilde{E}_{\mm{T}} = g^* \mu_{\mm{B}} \tilde{B} $.
This effective Hamiltonian differs from previous
derivations~\cite{ribeiro2009,ribeiro2010} in the coupling $f(t) =
c(\varepsilon(t))\lambda$, which is time-dependent.  Here, $c(\varepsilon(t))$ and
$\lambda$ are given in Eqs.~(\ref{eq:cepsilon}) and (\ref{eq:effective_hf_coupling}).  As
it can be seen from the functional form of $f(t)$, the effective coupling strength
between the spin states depends on the charge state (Fig.~\ref{fig:coupling}).  This
result is rather natural since the matrix element between $\mm{S}$ and $\mm{T}_+$ goes to
zero when the detuning is such that $\mm{S}=\mm{S}(2,0)$ and $\mm{T}_+ = \mm{T}_+(1,1)$,
i.e., $\bra{\mm{S}}H_{\mm{HF}}^{\perp}\ket{\mm{T}_+} \to 0$ for $\varepsilon \gg u$. It
also implies that the physics described by Eq.~\eqref{eq:eff_ham} goes beyond standard
LZSM theory with a constant coupling.
However, since both Hamiltonians describe adiabatic passage through an anti-crossing, we
can assume that the dynamics are qualitatively similar. We can therefore expect that our
previous discussion about enhancement of adiabaticity based on LZSM physics remains
valid. Furthermore, due to the peculiar form of $f(t)$ (c.f. Fig.~\ref{fig:coupling}), we
can assume that it is possible to observe finite-time interferometry phenomena in close
vicinity of the anti-crossing.

\subsection{Master Equation}

In order to compare our theory with experimental measurements, it is not sufficient to
solve the Hamiltonian dynamics provided by Eq.~\eqref{eq:eff_ham}, because some important
phenomena that can influence the outcome of the experiment are not taken into account.
Among these are spin relaxation due to phonon-assisted hyperfine
interaction~\cite{abalmassov2004} and charge fluctuations that lead to dephasing of the
qubit states~\cite{hu2006}. These phenomena can be taken into account in a quantum master
equation formalism for the density matrix. In the Lindblad
formalism~\cite{lindblad1976,breuer}, the time evolution
can be expressed as
\begin{equation}
\dot{\rho} = -\frac{\ui}{\hbar}\left[H,\rho\right] +
\frac{1}{2}\sum_{j=1}^{N^2 -1} \left(\left[L_j \rho,
L_j^{\dag}\right] + \left[L_j, \rho
L_j^{\dag}\right]\right).
\label{eq:mastereqL}
\end{equation}
The operators $L_j$ are called Lindblad operators~\cite{lindblad1976,breuer} and they
describe the dissipative effect of the environment on the system in the Born-Markov
approximation, which consists of two assumptions that lead to Eq.~\eqref{eq:mastereqL}.
The Born approximation supposes a weak coupling between the system and the bath, while
the Markov approximation consists in neglecting any type of memory effects of the bath
during the system evolution. To fully describe the effect of the environment, one needs
$N^2 -1$ operators where $N$ is the dimension of the system's Hilbert
space~\cite{lindblad1976,breuer}. For the case of a two-level system, the Lindblad
operators are $L_1 = \sqrt{\Gamma_-} \sigma_-$, $L_2 = \sqrt{\Gamma_+} \sigma_+$, and
$L_3 = \sqrt{\Gamma_{\varphi}} \sigma_z$, where $\sigma_-$ and $\sigma_+$ are spin ladder
operators and $\sigma_z$ is the $z$ Pauli matrix. They respectively describe relaxation
from the excited state to the ground state with rate $\Gamma_-$, relaxation from the
ground state to the excited state with rate $\Gamma_+$, and pure dephasing with rate
$\Gamma_{\varphi}$.

By substituting Eq.~\eqref{eq:eff_ham} into Eq.~\eqref{eq:mastereqL} and using the
expression of the Lindblad operators, the first order differential equation for the
$\mm{S}$-$\mm{T}_+$ density matrix can be written as
\begin{widetext}
\begin{equation}
\begin{pmatrix}
\dot{\rho}_{11}\\
\dot{\rho}_{12}\\
\dot{\rho}_{21}\\
\dot{\rho}_{22}
\end{pmatrix}
=
\begin{pmatrix}
	-\Gamma_+ &  \frac{\ui}{\hbar} f(t) & -\frac{\ui}{\hbar} f(t) & \Gamma_- \\
	\frac{\ui}{\hbar} f(t) & -\frac{\ui}{\hbar}
	(E_{\mm{S}}(t)-\tilde{E}_{\mm{T}}) - \frac{1}{2}\left(\Gamma_+ + \Gamma_- +
	4\Gamma_{\varphi}\right) & 0 & -\frac{\ui}{\hbar} f(t) \\
	-\frac{\ui}{\hbar} f(t) & 0 & \frac{\ui}{\hbar} (E_{\mm{S}}(t) -\tilde{E}_{\mm{T}}) - \frac{1}{2}\left(\Gamma_+ + \Gamma_- +
	4\Gamma_{\varphi}\right) & \frac{\ui}{\hbar} f(t)\\
	\Gamma_+ & -\frac{\ui}{\hbar} f(t) & \frac{\ui}{\hbar} f(t) & -\Gamma_-
\end{pmatrix}
\begin{pmatrix}
\rho_{11}\\
\rho_{12}\\
\rho_{21}\\
\rho_{22}
\end{pmatrix}.
\label{eq:mast}
\end{equation}
\end{widetext}
This system of four coupled ordinary differential complex equations can be reduced to a system of
three coupled ordinary differential equations (Bloch equations) by introducing new real
variables defined by
\begin{equation}
\begin{aligned}
x &= \rho_{12} + \rho_{21},\\
y &= \ui\left(\rho_{12} - \rho_{21}\right),\\
z &= \rho_{11} - \rho_{22}.
\end{aligned}
\label{eq:changetobloch}
\end{equation}
This set of variables is completed by the conservation of probability condition,
\begin{equation}
\rho_{11} + \rho_{22} = 1.
\label{eq:consprob}
\end{equation}
Substituting Eqs.~\eqref{eq:changetobloch}~and~\eqref{eq:consprob} into
Eq.~\eqref{eq:mast}, one finds the system of ordinary differential equations for the new
variables,
\begin{empheq}{align}
\dot{x} &= - \frac{E_{\mm{S}} - \tilde{E}_{\mm{T}}}{\hbar} y -
\frac{1}{2}\left(\Gamma_+ + \Gamma_-\right) x - 2 \Gamma_{\varphi} x,\\
\dot{y} &= \frac{E_{\mm{S}} - \tilde{E}_{\mm{T}}}{\hbar} x - 2 \frac{f}{\hbar} z -
\frac{1}{2}\left(\Gamma_+ + \Gamma_- \right) y - 2 \Gamma_{\varphi} y,\\
\dot{z} &= 2\frac{f}{\hbar} y - \left(\Gamma_+ + \Gamma_- \right) z +
\Gamma_- - \Gamma_+.
\label{eq:blocheq1}
\end{empheq}
\begin{figure}
\includegraphics[width=0.48\textwidth]{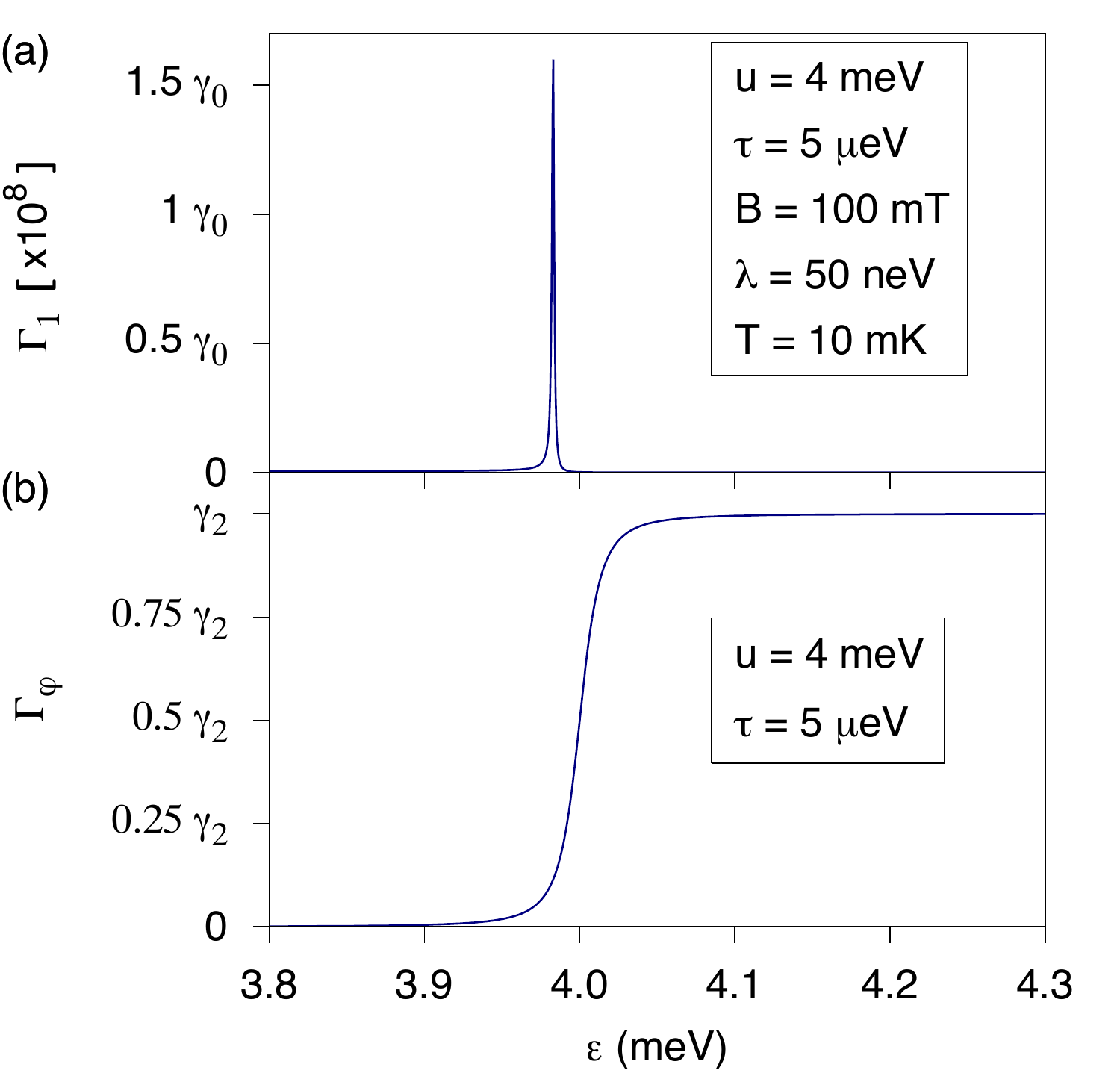}
\caption{(color online) (a) Relaxation rate $\Gamma_1 (\varepsilon)$ obtained for
$B=100\,\mm{mT}$. Here, we consider hyperfine mediated flip-flops of the electron
spin accompanied by emission or absorption of phonons. The relaxation rate is
maximal for $\varepsilon=\varepsilon_{\mm{c}}$ where $\abs{E^d_{\mm{S}}
(\varepsilon) - E^d_{\mm{T}}} \ll k_{\mm{B}} T$. This results in a strong mixing
of the spin states due to thermal fluctuations. (b) Charge noise induced
dephasing rate $\Gamma_2(\varepsilon)$. The dephasing rate is assumed to have a
functional dependence proportional to $1 - c(\varepsilon)^{2}$ to ensure that
dephasing related to charge noise is weaker when the system is in a $(1,1)$ charge
configuration.}
\label{fig:rates}
\end{figure}

Here, we assume that relaxation occurs through phonon-assisted hyperfine interaction.
Since we are dealing with small energy transfers, we consider only piezo
phonons~\cite{khaetskii2001}. The Hamiltonian describing the coupling between the
logical qubit states (i.e. the instantaneous energy eigenstates of
Eq.~\eqref{eq:eff_ham}) and the phonons is given by $H_{\mm{qp}} = \sigma_z \otimes
U_{\mm{ph}}$, with
\begin{equation}
	U_{\mm{ph}} (\rv, t) = \sum_{\nu,\qvs} \sqrt{\frac{\hbar}{2
		\rho \omega_{\nu,\qvs}}} A_{\nu,\qvs}\left[ \ue^{\ui(\qvs\cdot\rvs
		- \omega_{\nu,\qvss}t)} b_{\nu,\qvs}^{\dagger} + h.c.  \right],
	\label{eq:phononH}
\end{equation}
and $\sigma_z$ is the pseudospin $z$-Pauli matrix for $\ket{\mm{S}}$ and $\ket{\mm{T}}$.
Here, $b_{\nu,\qvs}^{\dagger}$ ($b_{\nu,\qvs}$) creates (annihilates) a phonon with
polarization $\nu$ and wave vector $\qv$. $A_{\nu,\qvs}$ is the effective piezoelectric
modulus, which depends only on the direction of $\qv$.

The rates $\Gamma_{\mm{+}}$ and $\Gamma_{\mm{-}}$ can be derived using Redfield
theory~\cite{redfield1957,chirolli2008}. We have
\begin{equation}
	\Gamma_{\pm} = 4 \abs{\bra{0}\sigma_z\ket{1}}^2 J_{\pm}(\omega),
	\label{eq:ratesredf}
\end{equation}
where $\ket{0}$ and $\ket{1}$ are the qubit states (i.e. in our case the eigenstates of
Eq.~\eqref{eq:eff_ham}), $\omega$ the angular frequency defined by the difference in
energy of the qubit states, and the spectral densities $J_{\pm} (\omega)$ are defined by
\begin{equation}
	J_{\pm} (\omega) = \int_{0}^{\infty}\id{t} \ue^{\mps \ui \omega t}\left\langle U_{\mm{ph}} (\rv, 0)
	U_{\mm{ph}} (\rv, t) \right\rangle.
	\label{eq:spectral_density}
\end{equation}
Here, $\langle \dots \rangle = \mm{Tr}\left[\dots \rho_{\mm{ph}}\right]$ with
$\rho_{\mm{ph}}$ the density matrix of the phonon bath at thermal equilibrium.

The evaluation of Eq.~\eqref{eq:spectral_density} leads to
\begin{equation}
	J_+ (\omega) =
	\sum_{\nu} \frac{3 \langle A^2 \rangle}{4 \pi^2 \rho c^3_{\nu}
	\hbar} \omega n(\omega),
\label{eq:j+}
\end{equation}
and
\begin{equation}
	J_- (\omega) =
	\sum_{\nu} \frac{3 \langle A^2 \rangle}{4 \pi^2 \rho c^3_{\nu} \hbar}
	\omega (1+n(\omega)),
\label{eq:j-}
\end{equation}
with $c_{\nu}$ the sound velocity and $n(\omega)$ the Bose-Einstein occupation
number, $n(\omega)= (\exp(\beta \hbar \omega)-1)^{-1}$, where $\beta = 1/k_{\mm{B}} T$.
$k_{\mm{B}}$ is the Boltzman's constant and $T$ is the phonon bath temperature, and
$\langle A^2 \rangle$ denotes an average piezoelectric modulus. In the
following, we denote the result of the sum over $\nu$ by $\gamma_0$.

Since $\Gamma_{\mm{+}}$ and $\Gamma_{\mm{-}}$ can be related to each other by
considering the limiting case of thermal equilibrium, Eqs.~\eqref{eq:blocheq1} can be
simplified to include only two independent rates. If the system reaches thermal
equilibrium then the detailed balance equation $\rho_{11}^{\mm{th}}\Gamma_{\mm{+}} =
\rho_{22}^{\mm{th}}\Gamma_{\mm{-}}$ holds. Moreover, the populations are given by the
canonical ensemble, $\rho_{ii}^{\mm{th}} = \exp(-\beta E_i)/Z$, with $Z$
the partition function. This yields
\begin{equation}
\frac{\Gamma_{\mm{+}}}{\Gamma_{\mm{-}}} =
\frac{\rho_{22}^{\mm{th}}}{\rho_{11}^{\mm{th}}} =
\ue^{-\beta \hbar \omega},
\label{eq:rateth}
\end{equation}
where we used $\hbar \omega = \sqrt{(E_{\mm{S}}(\varepsilon)-\tilde{E}_{\mm{T}})^2 + 4
f^2(\varepsilon)}$, which is the energy difference between the eigenstates of
Eq.~\eqref{eq:eff_ham}.

Combining the results of Eqs.~\eqref{eq:ratesredf},~\eqref{eq:j+},~\eqref{eq:j-},
and~\eqref{eq:rateth}, we find that
\begin{equation}
\Gamma_1 (\varepsilon) = \Gamma_{\mm{+}} + \Gamma_{\mm{-}} = \gamma_0
\frac{f^2(\varepsilon)}{\hbar^2 \omega} \coth\left( \frac{
\hbar \omega}{2 k_{\mm{B}} T}\right).
\label{eq:relaxationrate}
\end{equation}
This function is plotted against $\varepsilon$ in Fig.~\ref{fig:rates}(a), and has a peak
at $\varepsilon = \varepsilon_{\mm{c}}$ where $\hbar \omega \ll k_{\mm{B}} T$ resulting
in a strong mixing of the states due to thermal fluctuations.

Pure dephasing can originate from orbital effects. In our current description of the
problem, we have neglected that the wave functions of the singlet $\ket{\mm{S}(2,0)}$ and
triplet $\ket{\mm{T}_+ (1,1)}$ couple differently to the background charge environment
due to their different charge configurations. Thus, a superposition state of the form
$\ket{\psi} = \alpha \ket{\mm{S}(2,0)} + \beta \ket{\mm{T}_+ (1,1)}$ is sensitive to
background charge fluctuations (charge noise), which leads to dephasing of the state
$\ket{\psi}$~\cite{coish2005,hu2006}. If the qubit is in a superposition with same charge
state, $\ket{\varphi} = \alpha \ket{\mm{S}(1,1)} + \beta\ket{\mm{T}_+ (1,1)}$, then
the effects of charge noise are assumed to become weaker. Therefore, we choose to write
the charge induced dephasing as
\begin{equation}
\Gamma_{\varphi} (\varepsilon) = \gamma_2 (1 - \abs{c(\varepsilon)}^2),
\label{eq:puredephasing}
\end{equation}
where $\gamma_2$ is the charge noise rate.  In Fig.~\ref{fig:rates}, we plot the rates
$\Gamma_1 (\varepsilon)$ and $\Gamma_{\varphi} (\varepsilon)$.

Finally, the Bloch equations, written in a matrix form, and describing the dynamics
around the $\mm{S}$-$\mm{T}_+$ anti-crossing are
\begin{widetext}
\begin{equation}
\hbar
\begin{pmatrix}
\dot{x}\\
\dot{y}\\
\dot{z}
\end{pmatrix}
=
\begin{pmatrix}
-\frac{1}{2}\hbar(\Gamma_1 (\varepsilon (t)) + 4 \Gamma_{\varphi} (\varepsilon (t))) & -
E_{\mm{S}} (\varepsilon (t)) + \tilde{E}_{\mm{T}} & 0 \\
E_{\mm{S}} (\varepsilon (t))  - \tilde{E}_{\mm{T}} & -\frac{1}{2}\hbar(\Gamma_1
(\varepsilon (t)) + 2 \Gamma_{\varphi} (\varepsilon (t))) & -2 \lambda
c(\varepsilon(t))\\
0 & 2 \lambda c(\varepsilon(t)) & -\hbar\Gamma_1 (\varepsilon (t))
\end{pmatrix}
\begin{pmatrix}
x\\
y\\
z
\end{pmatrix}
\pm \hbar
\begin{pmatrix}
0\\
0\\
\gamma_1
\end{pmatrix}.
\label{eq:blocheqmatrix}
\end{equation}
\end{widetext}
The different signs in front of the inhomogeneous term come from the fact that the
states $\ket{\mm{S}}$ and $\ket{\mm{T}}$ exchange their roles as ground and excited
state of the system at $\varepsilon = \varepsilon_{\mm{c}}$. The spontaneous relaxation
rate $\gamma_1$ is given by
\begin{equation}
	\gamma_1 = \gamma_0 \frac{f^2(\varepsilon)}{\hbar^2 \omega}.
	\label{eq:spontaneous_rate}
\end{equation}

\section{Results}
\label{sec:results}

In the following we first present a comparison between experimental and theory results
obtained with experimental pulse profiles, which were measured at the output port of the
waveform generator, and experimentally determined $E_{\mm{S}} (\varepsilon)$ and
$c(\varepsilon)$. We then show further theory results for which we study the effect of
charge dynamics and phonon-mediated hyperfine relaxation.  For this purpose, we have
solved Eq.~\eqref{eq:blocheqmatrix} for different $\gamma_0$'s and $\gamma_2$'s.

The singlet return probability is obtained by solving Eq.~\eqref{eq:blocheqmatrix} for
different realizations of $\lambda$. The average value of $P_{\mm{S}}$ is then calculated
according to
\begin{equation}
P_{\mm{S}} = \frac{1}{N}\sum_{i=1}^{N}\frac{1}{2}\left(1 + z^{(i)}\right),
\label{eq:rsingletbloch}
\end{equation}
where $z^{(i)}$ is the $i$th solution of $z(t)$ in Eq.~(\ref{eq:blocheqmatrix}).

\subsection{Experimental observation of finite-time effects}

\begin{figure}
\includegraphics[width=0.48\textwidth]{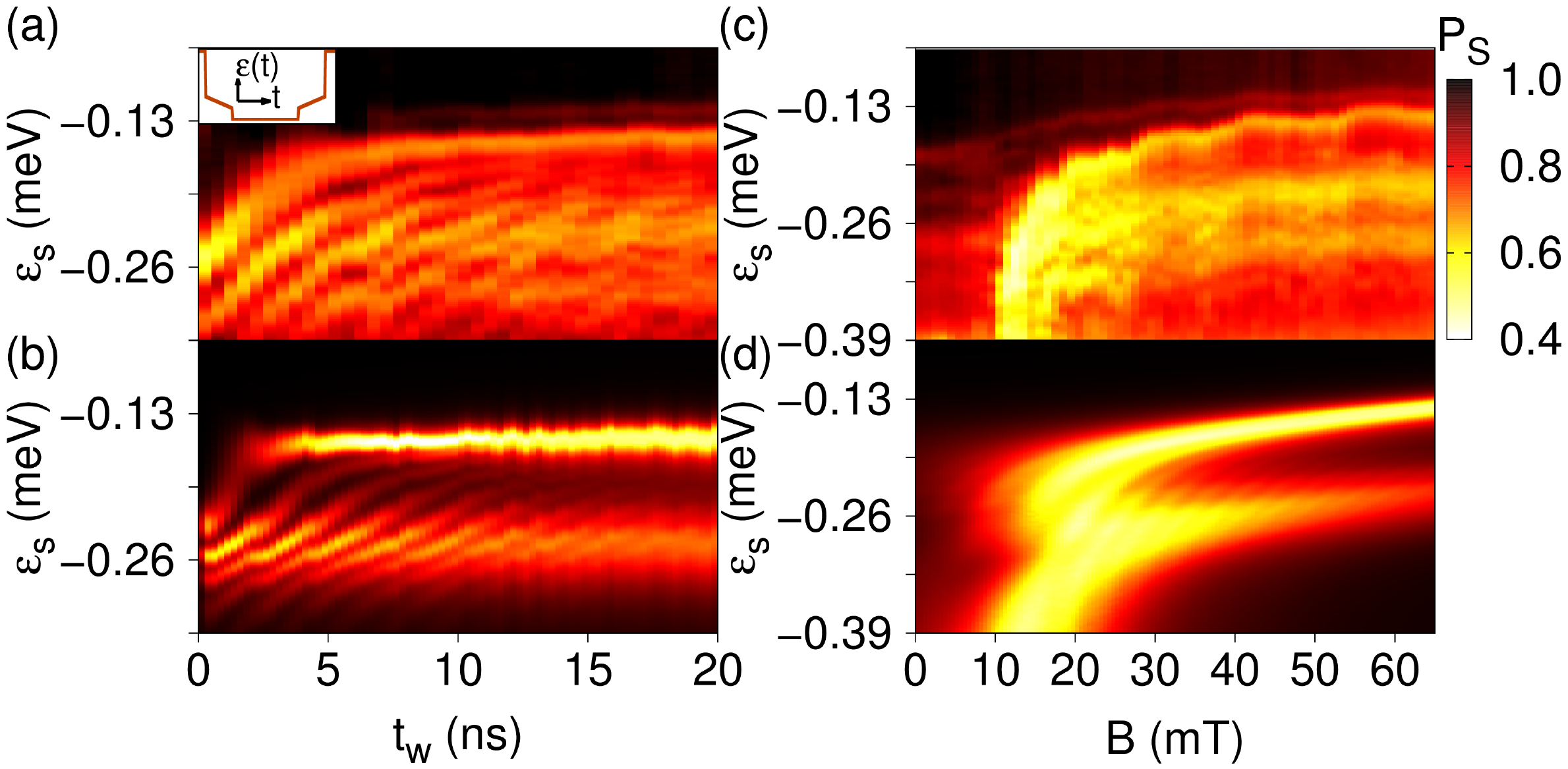}
\caption{(color online) (a) Measurements of $P_{\mm{S}}$ as a function of $t_{\mm{w}}$
and $\varepsilon_{\mm{s}}$ for $A_{1\mm{f}} = -0.26 \mm{meV}$, $t_{\mm{slow}} = 4\,\mm{ns}$,
and $B=55\,\mm{mT}$. ``Double hat'' pulses allow the observation of non-adiabatic
transitions when the system is driven slowly to close proximity of the anti-crossing. (b)
Theoretical predictions obtained using a pulse profile obtained at the output of the
waveform generator and parameters from (a). (c) $P_{\mm{S}}$ plotted as a function of $B$
and $\varepsilon_{\mm{s}}$ for $t_{\rm w} = 20\,\mm{ns}$ reveals the spin-funnel.
Parameters are $A_{1\mm{f}} = -0.26\,\mm{meV}$ and $t_{\mm{slow}} = 4\,\mm{ns}$.  (d)
Theoretical calculations for the same parameters as in (c).}
\label{fig:dhpulse1}
\end{figure}

We consider a ``double hat'' detuning pulse whose profile is shown in the inset of
Fig.~\ref{fig:dhpulse1}(a). The leading edge has a rise-time of $0.1\,\mm{ns}$ and an
amplitude $A_{1\mm{f}}$, which is followed by a slow ramp with rise-time $t_{\mm{slow}}$
and amplitude $A_{\mm{s}}=-0.065\,\mm{meV}$. A $0.1\,\mm{ns}$ rise-time pulse shifts the
detuning to its maximal value of $-0.39\,\mm{meV}$, where the detuning is held constant
for a time interval $t_{\mm{w}}$. The trailing edge of the pulse is simply the reverse of
the leading edge. The conversion between gate voltage and energy is performed using the
measured lever-arm $\sim 0.13\,\mm{meV}/\mm{mV}$~\cite{dicarlo2004,petta2004}. In
Fig.~\ref{fig:dhpulse1}, we compare experiment and theory for $t_{\mm{slow}} =
4\,\mm{ns}$, $A_{1\mm{f}} = -0.26\,\mm{meV}$.

$P_{\mm{S}}$ is plotted as a function of $t_{\mm{w}}$ and $\varepsilon_{\mm{s}}$ for
$B=55\,\mm{mT}$ in Fig.~\ref{fig:dhpulse1}(a) and (b). A ``spin-funnel'' that is obtained
using the spectroscopy method developed in Ref.~\onlinecite{petta2005} with a waiting
time $t_{\mm{w}}=20\,\mm{ns}$ is shown in Fig.~\ref{fig:dhpulse1}(c) along with the
corresponding theory plot in Fig.~\ref{fig:dhpulse1}(d). Since the experimental cycles
have a short period of $5\,\mm{\mu s}$, there is a build up of nuclear polarization that
generates a gradient field~\cite{foletti2009}. To take this into account in our model, we
add a mean to the nuclear field distributions Eq.~\eqref{eq:nuclfielddist} . It is
sufficient to consider only a mean $\xi_1^x$ for $B_{\mm{n},1}^x$ because only the
magnitude of the gradient field plays a role in the dynamics. This can be easily
understood by considering a rotation about the $z$-axis of the coordinate system that
brings the $x$-axis to coincide with the direction of $\Bv_{\mm{n},2}^{\perp} -
\Bv_{\mm{n},1}^{\perp}$.  Parameters used in the theory panels are $\delta = 1\,\mm{mT}$,
$\xi_1^x = 10\,\mm{mT}$, $\gamma_0 = 10^{-2}$ and $\gamma_2 = 10^8\,\mm{s}^{-1}$.

Both the experimental results and numerical simulations show enhanced interference
visibility within the region between $\varepsilon_{\mm{s}}$ $\sim -0.19\,\mm{meV}$ and
$\sim -0.29\,\mm{meV}$. This region should correspond to values of $P_{\mm{S}}$ determined
by the slow rise-time component of the pulse, as demonstrated in
Ref.~\onlinecite{ribeiro2012}. The position of the anti-crossing is located at $\sim
-0.14\,\mm{meV}$ from the data shown in Fig.~\ref{fig:dhpulse1}(a) and (c). Moreover, since
$A_{1\mm{f}} = -0.26\,\mm{meV}$, $A_{\mm{s}}=-0.065\,\mm{meV}$, and the maximal pulse amplitude
is $-0.39\,\mm{meV}$, the high contrast region should be located between $-0.21\,\mm{meV}$ and
$-0.27\,\mm{meV}$, in good agreement with our results.

As discussed in Sec.~\ref{sec:adiabatic}B, the visibility of the oscillation pattern
contained between $\varepsilon_{\mm{s}}$ $\sim -0.14\,\mm{meV}$ and $\sim -0.21\,\mm{meV}$
cannot result from the second fast rise-time portion of the pulse.  Although it drives
the system through the anti-crossing, the level velocity is too high to allow for
large magnitude non-adiabatic transition probabilities. Thus, as we showed in
Figs.~\ref{fig:dhlzsm} and \ref{fig:comp_finite-time} by considering a finite-time LZSM
model, we are able to observe a non-adiabatic transition event due to a slow level
velocity pulse that brings the system to close vicinity of the anti-crossing, but without
driving it through.

To ensure that we are observing finite-time effects, we consider a second ``double hat''
pulse that has a different $A_{1\mm{f}}$, which changes the relative starting and
stopping position of the slow rise-time component of the pulse. As a consequence the
relative propagation times $t_{\mm{i}}$ and $t_{\mm{f}}$, which are defined by setting
$t=0$ at the anti-crossing, are also modified.

\begin{figure}
\includegraphics[width=0.48\textwidth]{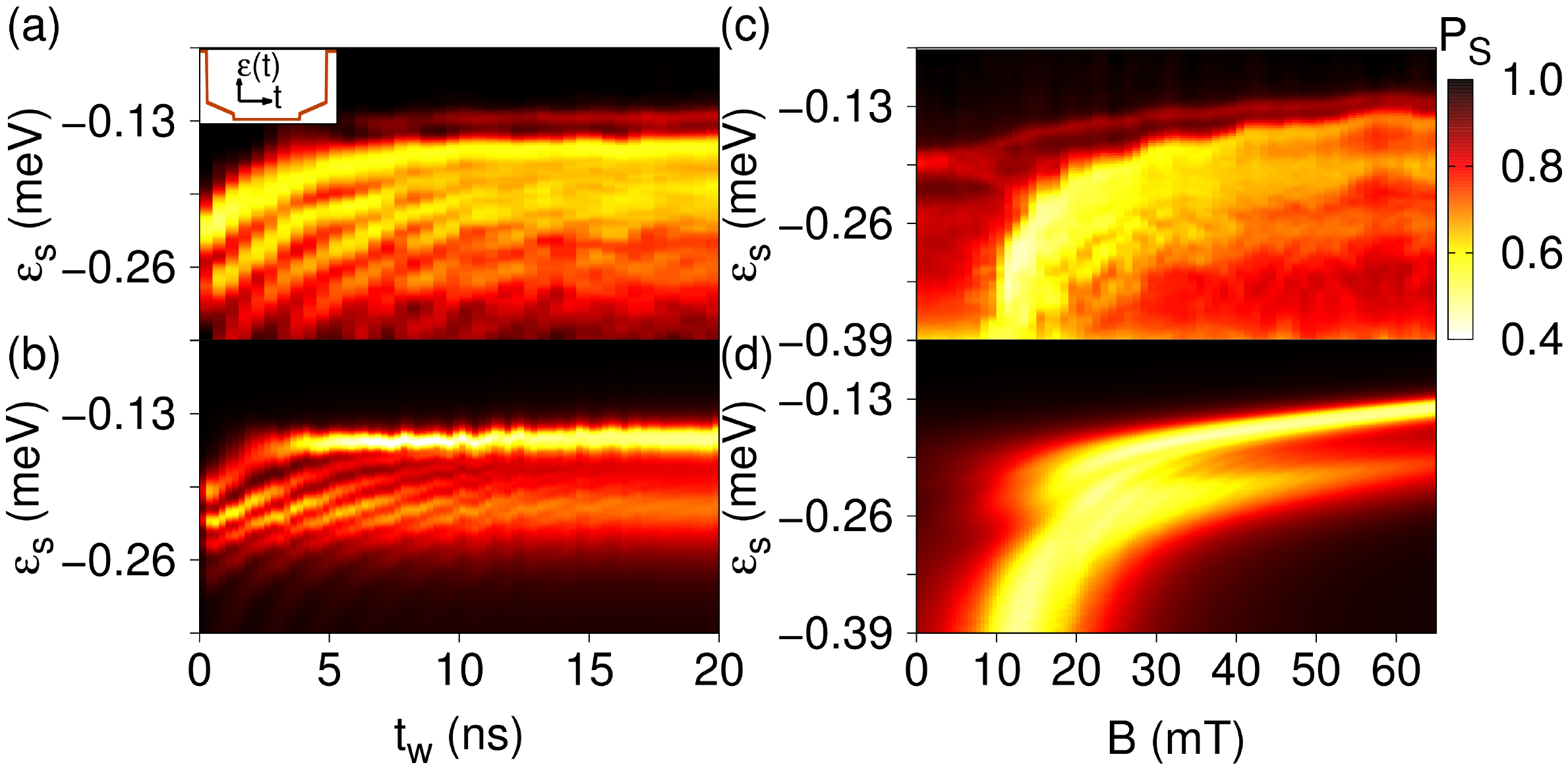}
\caption{(color online) Experimentally, (a) and (c), and theoretically, (b) and (d),
obtained LZSM interference patterns and spin-funnels. Here $A_{1\mm{f}} =
-0.29\,\mm{meV}$, $t_{\mm{slow}} = 4\,\mm{ns}$, and $B=55\,\mm{mT}$. The waiting
time for the spin-funnel is $t_{\rm w}$ = $20\,\mm{ns}$. The interference
patterns differ from results presented in Fig.~\ref{fig:dhpulse1}, indicating
that the evolution of the system is not only sensitive to the level velocity, but also
the time at which the change in level velocity happens.}
\label{fig:dhpulse2}
\end{figure}

We consider a second pulse with $A_{1\mm{f}}=-0.29\,\mm{meV}$, for which the results are
presented in Fig.~\ref{fig:dhpulse2}.
First, we observe that the high-contrast region is shifted towards more positive
detunings, as expected. Second, we notice the overall difference between the
interference pattern in Figs.~\ref{fig:dhpulse1} and~\ref{fig:dhpulse2}, panels (a) and
(b). This dissimilarity can only be explained by different phase accumulation due to
distinct $t_{\mm{i}}$ and $t_{\mm{f}}$. In the usual scattering description of LZSM
interferometry, the transition probability only depends on the level velocity at the
anti-crossing and on the coupling strength. But here, these two quantities have remained
unchanged. These results show the importance of using finite-time models to describe
adiabatic passage experiments.

We also anticipate, for LZSM driven qubits, the possibility to manipulate the states by
keeping a constant driving and instead change the relative starting and stopping position
of the detuning pulse.

\subsection{Effects of Phonon-Mediated Hyperfine Relaxation and Charge Induced Dephasing}

\begin{figure}
\includegraphics[width=0.48\textwidth]{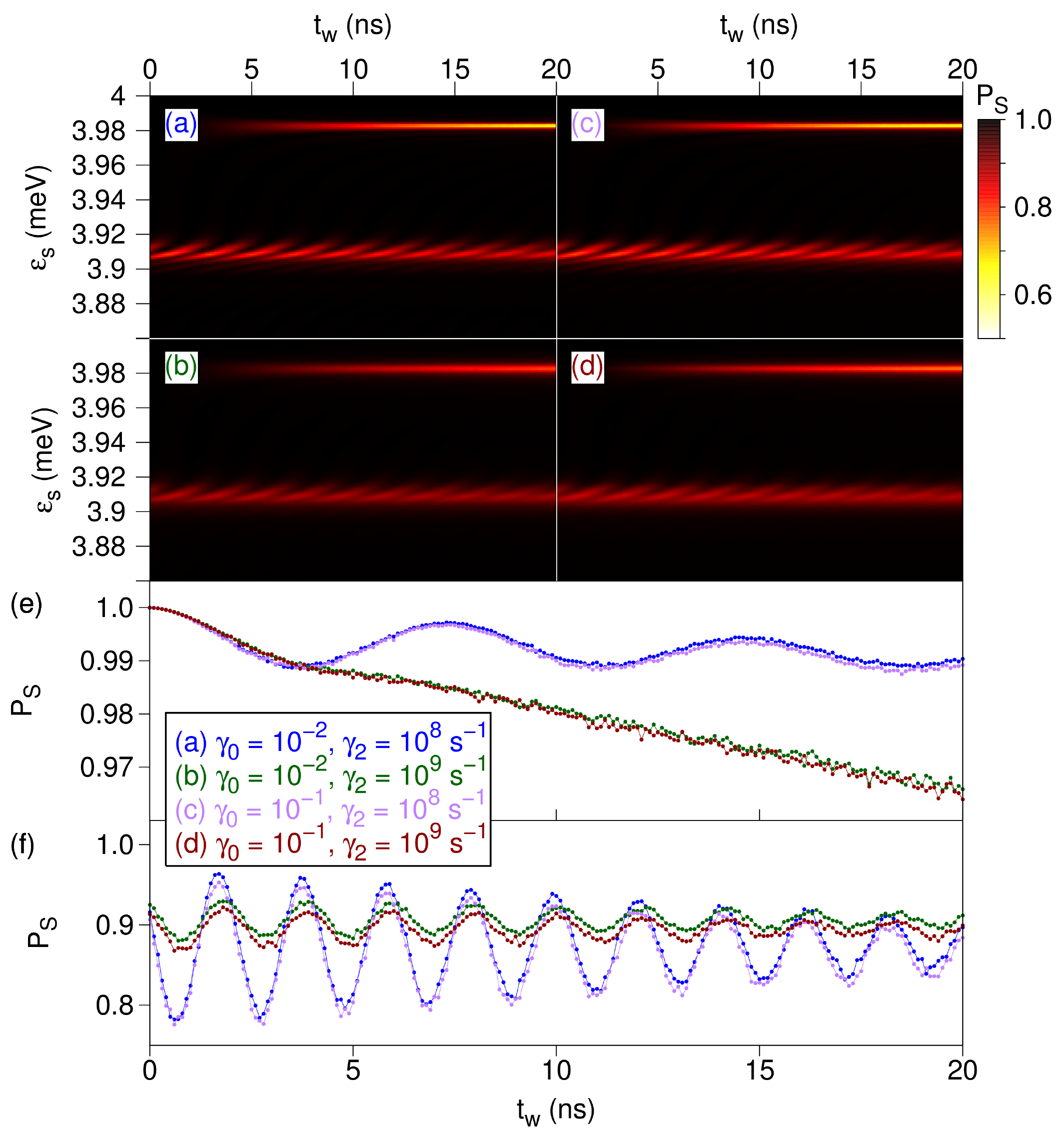}
\caption{(color online) Singlet return probability $P_{\mm{S}}$ as a function of the
waiting time $t_{\mm{w}}$ and final position $\varepsilon_{\mm{s}}$ obtained for a
``double hat'' detuning pulse. Pulse details are given in the text. Here, we
have set $B=100\,\mm{mT}$, $u=4\,\mm{meV}$, and $\tau=5\,\mm{\mu eV}$ for all plots. The
parameters $\gamma_0$ and $\gamma_2$ are (a) $\gamma_0 = 10^{-2}$, $\gamma_2 =
10^8\,\mm{s^{-1}}$, (b) $\gamma_0 = 10^{-2}$, $\gamma_2 = 10^9\,\mm{s^{-1}}$,
(c) $\gamma_0 = 10^{-1}$, $\gamma_2 = 10^8\,\mm{s^{-1}}$, and (d) $\gamma_0 =
10^{-1}\,\mm{s^{-1}}$, $\gamma_2 = 10^9\,\mm{s^{-1}}$. Figures (e) and (f) show cuts for
case (a) in blue, (b) in dark green, (c) in purple, and (d) in dark red. The cuts are
respectively taken before the anti-crossing at $\varepsilon_{\mm{s}} =
3.987\,\mm{meV}$ and after at $\varepsilon_{\mm{s}} = 3.908\,\mm{meV}$.}
\label{fig:resdh}
\end{figure}

In this section, we consider a detuning pulse $\varepsilon (t)$ with an amplitude of
$\varepsilon_{\mm{s}}
- \varepsilon_{\mm{i}} = 0.2\,\mm{meV}$. The pulse reaches an amplitude of
$0.12\,\mm{meV}$ in $0.1\,\mm{ns}$, it is then slowed down until it reaches an amplitude
of $0.1325\,\mm{meV}$, and finally it is brought to its maximal amplitude in
$0.1\,\mm{ns}$. The rise time $t_{\mm{slow}}$ of the slow part of the pulse can be tuned
freely. Here we compute $P_{\mm{S}}$ for $t_{\mm{slow}} = 8\,\mm{ns}$.  In
Fig.~\ref{fig:resdh}(a)-(d), we present the singlet return probability as a function of
$\varepsilon_{\mm{s}}$ and waiting time $t_{\mm{w}}$ for $B=100\,\mm{mT}$,
$u=4\,\mm{meV}$, $\tau = 5\,\mm{\mu eV}$, and different values of $\gamma_0$ and
$\gamma_2$.
The interference fringes are characterized, for $\varepsilon_{\mm{s}} < \varepsilon_{\mm{c}}$, by
three distinct regions showing an alternate oscillation amplitude for $P_{\mm{S}}$. The
darker regions coincide with detunings for which the passage through the anti-crossing
happened during one of the fast rise-times of $\varepsilon (t)$. Similarly, the bright
region between $\varepsilon_{\mm{s}}\simeq ~3.900-3.915\,\mm{meV}$ coincides with a passage
through the anti-crossing with the slow rise-time portion of the detuning pulse.
Experimentally measured interference patterns exhibit identical behavior.

Our results also clearly show coherent evolution of the qubit for $\varepsilon_{\mm{s}} >
\varepsilon_{\mm{c}}$, Fig.~\ref{fig:resdh}(e) [blue and purple trace, $\gamma_2 =
10^8\,\mm{s^{-1}}$]. This corresponds to the case where the system is not detuned
through the anti-crossing. It indicates that it is possible to design complex pulses that
can directly influence the competition between LZSM physics and charge noise. However,
this is conditional on the dephasing time scale associated with charge noise. We notice
that if $\gamma_2 = 10^9\,\mm{s^{-1}}$, then it is impossible to identify any coherent
behavior [green and red traces Fig. 7(e)].

Our results indicate that charge noise strongly affects the dynamics while spin
relaxation only has a moderate effect. Although this behavior can be identified when
comparing Fig.~\ref{fig:resdh}(a) with Fig.~\ref{fig:resdh}(b) and
Fig.~\ref{fig:resdh}(c), it is best seen in Fig.~\ref{fig:resdh}(f) when comparing
traces. An increase in the noise rate $\gamma_2$ leads to a substantial decrease of the
oscillation visibility (blue and green traces). On the other hand, the visibility is
only slightly diminished when relaxation is enhanced (blue and purple traces).

Here, charge noise has a drastic effect on spin dynamics. Since it leads to strong
dephasing of the logical qubit states when spin and charge degrees of freedom are
correlated, it results in a competition mechanism against LZSM
interferometry~\cite{ao1989,gefen1991,wubs2006,saito2007,nalbach2009,lehur2010}. Since
the hyperfine mediated anti-crossing is close to the charge anti-crossing for reasonable
values of $B$, charge noise also affects the dynamics during the passage through the
anti-crossing. As a result, the efficiency of the LZSM mechanism to create a coherent
superposition or to produce coherent interferences is hindered, and thus the optimal
state populations are not reached. This behavior can be identified in
Fig.~\ref{fig:resdh}(f) when comparing traces obtained with the same relaxation rate, but
different charge noise rates (i.e. blue with green trace and purple with red trace). We
clearly identify that the oscillations visibility is smaller for larger values of
$\gamma_2$.

Energy relaxation processes have a weaker influence on spin dynamics than dephasing due
to their dependence on the energy difference between the eigenstates of the system. These
become important only when the system is held in close vicinity of the anti-crossing,
where relaxation is maximum [c.f. Fig.~\ref{fig:rates}(a)].

These results suggest that inhomogeneous dephasing due to nuclear spin fluctuations is
not the only physical process that limits the coherence of a two-spin based qubit, but
also that charge noise plays a major role. This is an important result for future devices made
out of Si/SiGe which are reaching a maturity level comparable to
GaAs~\cite{maune2012,savage2012,wang2013}. Silicon based devices are very interesting candidates
for spin based quantum computing because the only stable isotope ($^{29}$Si) possessing a
nuclear spin ($I = 1/2$) has relatively low abundance, $\approx 5\%$. Thus, hyperfine
induced decoherence is weaker than in GaAs based nanostructures~\cite{maune2012}.

Consequently, we distinguish three time scales that govern the physics of partial spin
adiabatic passage in DQDs. There is the rise-time of the detuning pulse, the decoherence
time $T_2^*$, and $T_{\varphi}$ associated with charge noise induced dephasing. Ideally,
we would like to have rise-times shorter than $T_{\varphi}$ to preserve any spin
superposition state. But this would render adiabatic transitions unlikely and therefore
seriously hinder manipulation of the qubit. ``Double hat'' pulses partially solve the
problem, but for an even better result, it would be necessary to increase the coupling
between the qubit states. In GaAs double quantum dots, it is possible to prepare a
nuclear gradient field to enhance the hyperfine coupling~\cite{foletti2009}, with the
advantage of extending $T_2^*$.
In almost nuclear spin free systems (Si or C based DQDs), it is possible to use
micro-magnets to artificially induce a coupling between the $\mm{S}$-$\mm{T}_+$ qubit
states~\cite{pioroladriere2008}.

\section{Conclusions}

We have developed a master equation formalism to study the dynamics of a
$\mm{S}$-$\mm{T}_+$ qubit in the vicinity of the hyperfine mediated anti-crossing. In
comparison with previous theories that only included decoherence due to the hyperfine
interaction with nuclear spins, we also include phonon-mediated hyperfine spin relaxation
and spin dephasing due to charge noise. We have also derived a new effective two-level
spin-charge Hamiltonian. In addition to previous theories, we take into account the
charge degree of freedom. This property originates from the lowest energy singlet state
$\ket{\mm{S}}$ being a superposition of different charge configurations. Although the
effective coupling between the $\mm{S}$-$\mm{T}_+$ qubit states is time-dependent, its
form still allows to approximately reason in terms of LZSM physics.

With our formalism, we have compared results for different values of $\gamma_0$ and
$\gamma_2$. Our findings suggest that LZSM spin interferometry is largely inhibited by
charge dynamics, and thus charge coherence has to be treated on equal footing with spin
coherence. We have indeed demonstrated that the visibility of the coherent oscillations
of a spin qubit is sensitive to the time scale associated with charge induced spin
dephasing. We have also shown that this interplay between charge and spin can prevent the
observation of finite-time oscillations. Interestingly, this could lead to the
development of an experimental protocol to measure the time scale associated with charge
decoherence.

\section*{Acknowledgements}
The authors are grateful to H. Lu and A. C. Gossard for assistance with device
fabrication. H. R. and G. B. acknowledge funding from the DFG
within SPP 1285 and SFB 767. Research at Princeton was supported by the Sloan and Packard
Foundations and the National Science Foundation through the Princeton Center for Complex
Materials, DMR-0819860 and CAREER award, DMR-0846341.

\end{document}